\documentclass[twocolumn,showkeys,preprintnumbers,amsmath,amssymb,floatfix]{revtex4}
\usepackage{amsmath}
\usepackage{rotating}
\usepackage{framed}
\usepackage{hyperref}
\hypersetup{colorlinks,citecolor=blue}
\usepackage{graphicx}
\usepackage{dcolumn}
\usepackage{bm}
\usepackage{color}
\usepackage{float}

\begin{document}

\preprint{}

\title{Anisotropic solution for polytropic stars in $4D$ Einstein-Gauss-Bonnet gravity }

\author{Ksh. Newton Singh}
\email{ntnphy@gmail.com}
\address{Department of Physics, National Defence Academy,  Khadakwasla,\\
 Pune, Maharashtra-411023, India.}

\author{S. K. Maurya}
 \email{sunil@unizwa.edu.om}
\address{Department of Mathematical and Physical Sciences, College of Arts and Sciences, University of Nizwa, Nizwa, Sultanate of Oman.}
 
\author{Piyali Bhar}
 \email{piyalibhar@associates.iucaa.in}
\affiliation{Department of Mathematics,Government General Degree College, Singur, Hooghly, West Bengal 712 409, India.}

\author{Riju Nag}
 \email{rijunag@gmail.com}
\affiliation{Department of Mathematical and Physical Sciences, College of Arts and Sciences, University of Nizwa, Nizwa, Sultanate of Oman.}

\date{\today}
\begin{abstract}
In the present work we have investigated a new anisotropic solution for polytropic star in the framework of $4D$ Einstein-Gauss-Bonnet (EGB) gravity. The possibility of determining the masses and radii of compact stars which puts some limitations on equation of state (EoS) above the nuclear saturation density. For this purpose, the $4D$ EGB field equations are solved by taking a generalized polytropic equation of state (EoS) with Finch-Skea ansatz. The generalized solution for anisotropic model has been tested for different values of Gauss-Bonnet constant $\alpha$ which satisfies all the physical criteria including causality with static stability via mass vs central mass density ($M-\rho_c$), Bondi and Abreu criterion. The adiabatic index shows a minor influence of the GB coupling constant whereas the central and surface redshifts in the EGB gravity always remain lower than the GR. We present the possibility of fitting the mass and radius for some known compact star via $M-R$ curve which satisfies the recent gravitational wave observations from GW 170817 event.  
\end{abstract}

\maketitle


\section{Introduction}
Einstein's general theory of relativity has been a breakthrough theory and has played a pivotal role in understanding the nature of the universe. Despite its huge success, this theory has some drawbacks, such as it cannot explain the accelerated expansion of the universe. Also, this theory does not generalize well to dimensions other than the 4-dimensional framework. To address these issues, two distinct approaches have been incorporated. One of them is to change the matter part of Einstein's theory  which leads to the dark matter and dark energy hypothesis. Another approach is to modify the gravitational part of the Einstein-Hilbert (E-H) action and by this method, several modified gravity theories have emerged. From them, the higher derivative gravity theories have gained a lot of interest in the recent past as they have shown some potentiality in solving problems like the singularity problems of black holes. From them, Lovelock gravity \cite{Lovelock1,Lovelock2} is one of the notable ones. It is a generalized metric theory of gravity, for arbitrary $D-$dimensions which yields conserved second-order equations of motion. This is an effective way to generalize Einstein's gravity into higher dimensions and for $D=4$, we revert to Einstein's theory of gravity. For higher dimensional spacetime, along with the Einstein-Hilbert term and the cosmological constant, a Gauss-Bonnet (GB) term is allowed in Lovelock's action. When this GB term is added, it is being called Einstein-Gauss-Bonnet gravity. But while considering the $4D$ case, the GB term does not contribute as it becomes a topological invariant. So how does the 4-dimensional EGB gravity come into the picture? There is one specific approach that makes it possible. This methodology is known as regularization which was first used by Tomozawa \cite{Tomozawa} who made finite and one-loop quantum corrections to Einstein's gravity. Following a similar methodology, a dimension regularization of the GB equations was proposed by Glavan and Lin \cite{Glavan} and a $4D$ metric theory was obtained that can avoid Ostrogradsky instability. The methodology was constructed in $D-$dimensions and then with the rescaling of the coupling constant $\hat{\alpha} \to \alpha/(D-4)$ and considering the limit $D \to 4$, the GB term now shows non-negligible contribution to the gravitational dynamics, and thus the $4D-$EGB gravity works. This approach subsequently became popular among researchers investigating astrophysical solutions and their properties. In the context of realistic hadronic, the mass-radius relations were obtained by Doneva \& Yazadjiev \cite{Doneva1} using strange quark star equation of state (EoS) and have been studied in several contexts such as static and spherically symmetric black hole solutions and their physical properties \cite{g8,g9,g10,g11,g12,g13,g14}. Other problems including black holes having charge \cite{g15,g16}, black holes coupled with nonlinear electrodynamics and magnetic charge \cite{g17,g18,g19} have been studied in this context. 

The study of the compact stars has also attracted much attention to the researchers in the modified theories of gravity. So far, there is no exact EoS known that describes the internal structure of relativistic compact objects. Many approaches have been taken to model the compact stars, including considerations of isotropic fluid, anisotropic fluid, charged isotropic, and charged anisotropic fluid inside the compact objects. But, obtaining exact solutions for isotropic scenarios is not an easy task as compared to anisotropic solutions. Moreover, in extremely dense conditions, the pressures may actually bifurcate to radial and tangential components and that can lead to pressure anisotropy. It was Ruderman \cite{ruderman} who showed that in densities higher than $10^{15} g/cm^3$, the nuclear matter becomes anisotropic. Moreover, the positive anisotropy shows that the anisotropic force acts radially outwards and it helps in preventing gravitational collapse. The pressure anisotropy has been analyzed by a number of researchers in the context of compact stars and several of them can be found in the following references \cite{Mak,Kipp,Soko,Herrera1997,Hrr1,Hrr2,Hrr3,Hrr4,Sulaksono2019,Rizaldy2019,Maurya2019a}. Anisotropic quark stars in the context of $4D-$EGB gravity have been discussed by Banerjee and his collaborators \cite{banerjee,banerjee1}. On the other hand, some exact solutions in the context of compact stars and wormholes can be found in the following Refs\cite{4D1,4D2,4D3,4D4}. 

Extremely dense objects like neutron stars can have the presence of various exotic matter with a significant strangeness fraction such as quark matter, hyperon matter, and Bose-Einstein condensates of strange mesons in their interior. Also, some theories indicate that the presence of these exotic components makes the equation of state (EoS) of the compact stars soft and as a result, compact stars having a smaller radius and maximum mass can exist, in comparison to the stiffer EoS. \cite{lp}. But, highly massive neutron stars set rigid constraints while modeling the dense nuclear matter. These massive compact objects have great application in quantum chromodynamics (QCD), where, there is a phase transition inside the neutron star, between hadronic matter to deconfined quark. Irrespective of our understanding of QCD, as of now, the majority of the studies of the quark stars have been based on the MIT bag model (Chodos et al. \cite{1974a,1974b}; Peshier et al. \cite{2000}). According to this model, the quarks are inside the bag and are considered as free Fermi gas and thus it provides a mechanism of quark confinement. But, the MIT bag model has its own limitations. One of that is that even for the massless quarks, it violets the chiral symmetry. As a result, for a system with more complex structures and interacting quarks, this EoS is not sufficient. To address this issue, some researchers proposed modified versions of it, like color–flavor-locked (CFL) matter (Alford et al. \cite{1999}). In fact, for asymptotically large densities, this CFL matter can be a real ground state of QCD (Alford et al. \cite{1998}). However, as the phase of matter in the extremely dense condition is quite uncertain, for these specific conditions, Asbell \& Jaikumar \cite{2017} proposed a two-component model for quark stars which can produce stars as heavy as 2 solar masses. 

Based on the literature, the polytropic EoS $p_r=\gamma \rho^{1+\frac{1}{n}}$ has been widely used to study the properties of the compact objects \cite {Eos1,Eos2,Eos3,Eos4,Eos5,Eos6,Eos7,Eos8,Eos9,Eos14}. However, the generalized polytropic EoS $p_r=\gamma \rho^{1+\frac{1}{n}}+\beta \rho $ has been used first time to discuss various cosmological aspects of the universe \cite{Eos1c}. Later on, by taking negative indices in the case of a generalized polytropic EoS, Chavanis \cite{Eos2c} described the models in the context of the late universe. In this connection, Freitas and Goncalves \cite{Eos3c} applied a generalized polytropic EoS to study elemental quantum fluctuations and constructed a universe with constant density at the origin. According to the nature of EoS, we can categorize most of the EoS into two classes: (i) the normal EoS which has a vanishing pressure when the density goes to zero, (ii) self-bound equations of state in which pressure vanishes at a significant finite density. As we can see that the generalized EOS as mentioned above cannot describe the self-bound compact objects, therefore the said EoS was modified  to $p_r=\gamma \rho^{1+\frac{1}{n}}+\beta \rho +\chi $. This equation of state has been used by several authors to find the self-bound compact objects \cite{Eos10,Eos11,Eos12,Eos13}.

In our current work, with the consideration of $4D-$EGB gravity, we end up having three equations and five unknowns. To counter this,  we chose Finch-Skea $g_{rr}$ metric and a  polytropic EoS of the form $p_r = \gamma \rho^{1+\frac{1}{n}} +\beta \rho + \chi$ to close the system of equations completely, where $\gamma$, $\beta$, and $\chi$ are constant parameters and $p_r$ is the radial component of the pressure, while $n$ is polytropic constant.  

The present article is distributed in six sections, where the first section is the introduction. In the second section, the field equations were described in the context of $D-$ dimensional EGB gravity, and from there, the field equations of 4- dimensional EGB gravity were obtained. The third section consists of the analysis of the hybrid star solution. The boundary conditions were discussed in the fourth section using Glavan and Lin exterior solution. The physical analysis has been done in the fifth section and the last and final section, being the results and discussion.

\section{The field equations}\label{sec2}

The complete action in $D-$dimensional EGB gravity is \cite{Pedro}
\begin{eqnarray}
\mathcal{S}_{EGB} = {1 \over 16\pi} \int d^Dx \sqrt{-g} \left(\mathcal{R}-2\Lambda+\hat{\alpha} \mathcal{L}_{GB}\right)+\mathcal{S}_m. \label{b7}
\end{eqnarray}
The variation of \eqref{b7} with respect to the metric tensor gives the EGB field equation
\begin{eqnarray}
\mathcal{G}_{\mu \nu}+\Lambda g_{\mu \nu} =\hat{\alpha} \mathcal{H}_{\mu \nu} +8\pi \mathcal{T}_{\mu \nu} \label{b8}
\end{eqnarray}
where,
\begin{eqnarray}
\mathcal{G}_{\mu \nu} &=& \mathcal{R}_{\mu \nu} -{1 \over 2}~\mathcal{R}~g_{\mu \nu}~,\\
\mathcal{H}_{\mu \nu} &=& -2\Big(\mathcal{R} \mathcal{R}_{\mu \nu}-2\mathcal{R}_{\mu \lambda} \mathcal{R}^\lambda_\nu-2\mathcal{R}_{\mu \lambda \nu \rho} \mathcal{R}^{\lambda \rho}+ \nonumber \\
&& \mathcal{R}_{\mu \alpha \beta \gamma} \mathcal{R}^{\alpha \beta \gamma}_\nu -{1 \over 4}~g_{\mu \nu} \mathcal{L}_{GB}\Big), \label{b10}\\
\mathcal{T}_{\mu \nu} &=& -{2 \over \sqrt{-g}}~{\delta (\sqrt{-g} ~\mathcal{S}_m) \over \delta g^{\mu \nu}}.
\end{eqnarray}
Equation \eqref{b10} is anti-symmetric over five indices and hence must be vanishing for $D<5$. This can be seen through the trace of $\mathcal{H}_{\mu \nu}$ which can be written as \cite{Glavan}
\begin{eqnarray}
\hat{\alpha}\mathcal{H}^\mu_{~~\mu} = {\hat{\alpha}(D-4) \over 2} \mathcal{L}_{GB} \label{b12}
\end{eqnarray}
which is clearly vanishing under the limit $D\rightarrow 4$. Hence, the GB contribution in the field equation \eqref{b8} is nullified. However, if we adopt a re-scaling $\hat{\alpha} \rightarrow \alpha/(D-4)$ \cite{Glavan}, Eq. \eqref{b12} reduces to
\begin{eqnarray}
\hat{\alpha}\mathcal{H}^\mu_{~~\mu} = {\alpha \over 2} \mathcal{L}_{GB}.
\end{eqnarray}
Thus, the GB contribution is non-vanishing. Therefore, the final form of the field equation takes the form
\begin{eqnarray}
\mathcal{G}_{\mu \nu}+\Lambda g_{\mu \nu}-{\alpha \over D-4} \mathcal{H}_{\mu \nu}  = 8\pi \mathcal{T}_{\mu \nu}.
\end{eqnarray}
We have considered the above field equation avoiding the cosmological constant term.\\
With this re-scaling $\hat{\alpha} \rightarrow \alpha / (D-4)$, Ghosh \& Maharaj \cite{ghm} shown that by considering spacetime of curvature scale $\mathcal{K}$ which are maximally symmetric determined the variation of the Gauss-Bonnet contribution as
\begin{eqnarray}
{1 \over \sqrt{-g}}~g_{\mu \lambda}~{\delta \mathcal{L}_{GB} \over \delta g_{\nu \lambda}}={\alpha(D-2)(D-3) \over 2(D-1)}~\mathcal{K}^2~\delta^\nu_\mu
\end{eqnarray}
which is clearly non-vanishing at $D=4$.

To arrive at the reduced field equations we consider a spacetime in $D-$dimensions
\begin{eqnarray}
ds^2 &=&- e^\nu dt^2+e^\lambda dr^2+r^2 d\Omega^2_{D-2}. \label{e9} 
\end{eqnarray}
Here, $d\Omega^2_{D-2}$ represents the $D-2$ dimensional surface of a unit sphere. Further, assuming stress tensor for anisotropic fluid as
\begin{eqnarray}
\mathcal{T}_{\mu \nu} = (\rho+p_t)u_\mu u_\nu+p_t g_{\mu \nu}+(p_r-p_t)\chi_\nu\chi^\mu,
\end{eqnarray}
where all the symbols have their usual meanings.

Now, the field equations in the limit $D\rightarrow 4$ takes the form (for detailed derivation see \ref{appB})
\begin{small}
\begin{eqnarray}
&& \hspace{-0.4cm}  8\pi \rho = {e^{-\lambda} \lambda' \over r} \left[1+{2\alpha(1-e^{-\lambda}) \over r^2}\right]+{1-e^{-\lambda} \over r^2} \left[1-{\alpha(1-e^{-\lambda)} \over r^2} \right],~~~~ \label{e11}\\
&& \hspace{-0.4cm}  8\pi p_r= {e^{-\lambda} \nu' \over r} \left[1+{2\alpha(1-e^{-\lambda}) \over r^2}\right]-{1-e^{-\lambda} \over r^2} \left[1-{\alpha(1-e^{-\lambda)} \over r^2} \right],~~~~ \label{e12}\\
&& \hspace{-0.4cm}  8\pi p_t = {e^{-\lambda} \over 4} \bigg[(2\nu''+\nu'^2)\left\{1+{4\alpha(1-e^{-\lambda}) \over r^2} \right\} + {2(\nu'-\lambda') \over r}~\nonumber \\
&& \hspace{0.3cm} \times \left\{1-{2\alpha(1-e^{-\lambda}) \over r^2} \right\}-\lambda' \nu' \left\{1-{8\alpha \over r^2}+{12\alpha (1-e^{-\lambda}) \over r^2} \right\} \bigg] \nonumber \\
&& \hspace{0.5cm}-{2\alpha(1-e^{-\lambda})^2 \over r^4}. \label{e13}
\end{eqnarray}
\end{small}
In this paper, we have chosen four different values for $\alpha$ to discuss the physical analysis of the present model, where $\alpha=0$ corresponds to the GTR case.  In the work of Charmousis et al. \cite{Charmousis} both the positive and negative values of $\alpha$ were used to describe various physical features of the model. An upper bound of $\alpha$ was also obtained in this model. Recently Pretel and Benerjee \cite{ban1} proposed a model of a compact star in the framework of Einstein-Gauss-Bonnet theory in five-dimensional spacetime.  They have shown that in comparison to GR gravity, EGB gravity leads to more compact stars depending on the internal structure of the stars. The radius increases as the value of increases, while the gravitational mass decreases. Furthermore, for some positive values, greater maximum masses can be obtained; however, such configurations violate the causality condition.  They also considered both positive and negative values of $\alpha$. Clifton et al.\cite{clif} proposed the observational constraints  on the coupling parameter $\alpha$ for the regularized version of the $4D$ Einstein-Gauss-Bonnet theory of gravity.  They found an overall bound on the EGB coupling constant $0 \leq \alpha \leq 10^8\, m^2 $ in the context of  binary black hole systems. While Feng et al. \cite{feng} obtained the constraints on the coupling parameter $\alpha$ based gravitational waves (GWs) measured by GW170817 and GRB 1708 17A, which is $-7.78 \times 10^{-16} \leq \alpha\leq 3.33 \times 10^{-15}$.  In 5D EGB gravity, Bhar et al. \cite{bhar1,bhar2,bhar3} developed models of both charged and uncharged compact stars. Recently, Maurya et al. \cite{Maurya1} obtained an anisotropic model in $5D$ EGB gravity. They always select a positive alpha value for the Gauss-Bonnet term. Furthermore, Quark stars in 4-dimensional Einstein-Gauss-Bonnet gravity can be obtained recently in one of our previous paper \cite{4D1}. Inspired by all of these earlier works here we choose the values of $\alpha$.

\section{Anisotropic solution for polytropic star}
In this section, we focus on determining the closed form solution for the neutron star model. Since $4D-$EGB  field equations (\ref{e11})-(\ref{e13}) contain five unknowns $\{\rho, p_r, p_t, \nu, \lambda\}$, therefore we need two extra conditions to close the above system. For this purpose, we use a generalized polytropic equation of state (EoS) of the form,
\begin{eqnarray}
p_r &=& \gamma\, \rho^{1+1/n}+\beta \rho + \chi. \label{eq14}
\end{eqnarray}
Here $\gamma,~ \beta$ and $\chi$ are constant parameters with proper dimensions and $n$ denotes a polytropic index. Furthermore, the present polytropic EoS can represents a MIT bag EoS by taking $\gamma=0,~\beta=\frac{1}{4}$ and $\chi=-\frac{4}{3}\mathcal{B}_g$, where $\mathcal{B}_g$ is a bag constant. Therefore, $\gamma$ plays an important role to observe what kind of contribution is coming in the MIT bag model. To find the exact solution, we chose the polytropic index to be one i.e. $n=1$. The EoS \eqref{eq14} with $n=1$ has a quadratic contribution $\gamma \rho^2$, which usually expressed the neutron liquid in Bose-Einstein condensate form and the linear terms $\beta \rho+\chi$ come from the free quarks model of the famous MIT bag model, with specific values of $\beta=1/3$ and $\chi=-4\mathcal{B}_g/3$. Hence, these neutron stars are most likely ``{\it hybrid stars}''.\\

Then above EoS (\ref{eq14}) together with equations (\ref{e11}) and (\ref{e12}) give a non-linear differential equation of the form,
\begin{small}
\begin{eqnarray}
&& \hspace{-0.2cm} \alpha^2 \gamma (1 - e^{-\lambda})^2 (1 - e^{-\lambda} - 2 \lambda^{\prime} r e^{-\lambda})^2 +
 r^4 [r^2 (1 + \beta + \chi r^2 \nonumber\\&&\hspace{-0.2cm} - e^{-\lambda}  - \beta e^{-\lambda} - \nu^{\prime} r e^{-\lambda} + \lambda^{\prime} \beta r e^{-\lambda}) +
    \gamma (1 + (\lambda^{\prime} r-1) e^{-\lambda})^2] \nonumber\\&&\hspace{-0.2cm} -
 \alpha\, r^2  \times ( e^{-\lambda}-1)\, [2 \gamma \{ (2 + \lambda^{\prime} r) e^{-\lambda} -1 - (1 + \lambda^{\prime} r -2 \lambda^{\prime 2} r^2) \nonumber\\&&\hspace{-0.2cm} \times e^{-2\lambda}\} -
    r^2 (1 - e^{-\lambda} + 2 \nu^{\prime} r e^{-\lambda} + \beta (1 - e^{-\lambda} - 2 \lambda^{\prime} r e^{-\lambda}))]=0,~~~~~~~\label{eq15}
\end{eqnarray}
\end{small}
The above differential equation (\ref{eq15}) depends on the metric potentials $\nu$ and $\lambda$. Therefore, we chose the Finch-Skea ansatz for potential $\lambda$ as,
\begin{eqnarray}
\lambda=\ln(1+ar^2),
\end{eqnarray}
where $a$ is constant parameter with dimension $length^{-2}$. Now we solve the Eq.(\ref{eq15}) by plugging $\lambda$ and then obtain a closed form solution for other potential $\nu$ of the form,
\begin{small}
\begin{eqnarray}
&& \hspace{-0.2cm} \nu(r) = A+{1 \over 8} \bigg[\frac{a \gamma  (a \alpha -1)}{\pi  \left(a r^2+1\right)^2}-\frac{4 \alpha  a^2 \gamma }{3 \pi  \left(a r^2+1\right)^3}- \frac{a \gamma  (a \alpha +8)}{4 \left(\pi  a r^2+\pi \right)}\nonumber\\
&& \hspace{0.2cm} +\frac{16 \pi  \chi  \left(a r^2+1\right)^2}{a} +4 \left(a r^2+1\right) (1-16 \pi  \alpha  \chi +\beta)+\frac{(64 \pi  \beta -5 a \gamma ) }{8 \pi } \nonumber \\
&& \hspace{0.2cm} \log \left(a r^2+1\right)+\frac{a  \log \left(2 a \alpha +a r^2+1\right)}{8 \pi } \Big\{1024 \pi^2 \alpha ^2 \chi -96 \pi  \alpha  \nonumber \\
&& \hspace{0.2cm} (\beta +1)+9 \gamma \Big\}\bigg].
\end{eqnarray}
\end{small}

Now the expressions for the density, anisotropy and $p_t$ are
\begin{small}
\begin{eqnarray}
\rho(r) &=& \frac{a \left[a^2 \left(r^4-\alpha  r^2\right)+a \left(3 \alpha +4 r^2\right)+3\right]}{8 \pi  \left(a r^2+1\right)^3}\\
\Delta(r) &=& {1 \over 64} \Bigg[-\frac{8 a \beta  \left(a^2 \left(r^4-\alpha  r^2\right)+a \left(3 \alpha +4 r^2\right)+3\right)}{\pi  \left(a r^2+1\right)^3}- \nonumber \\
&& \hspace{-0.3cm} \frac{a^2 \gamma  \left(a^2 \left(r^4-\alpha  r^2\right)+a \left(3 \alpha +4 r^2\right)+3\right)^2}{\pi ^2 \left(a r^2+1\right)^6}-64 \chi \nonumber \\
&& \hspace{-0.3cm} \frac{\left(a r^2+1\right)^{-2}}{8 \pi} \Bigg\{-128 \alpha  a^2-a r^2 \left[f_1(r)+f_2(r)\right] \Big(1-\frac{12 \alpha }{a r^4+r^2} \nonumber \\
&& \hspace{-0.3cm} +\frac{4 \alpha }{r^2}\Big)  +\left[f_3(r)+f_2(r)\right] \left[a \left(r^2-2 \alpha \right)+1\right] +16 [a \left(4 \alpha +r^2\right)\nonumber \\
&& \hspace{-0.3cm} +1] \Bigg[\frac{r^2 \left(f_1(r)+f_2(r)\right){}^2}{1024}+f_4(r)+f_5(r)+f_6(r) \Bigg]\Bigg\} \Bigg] \\
p_t(r) &=& \Delta (r) +p_r (r).
\end{eqnarray}
\end{small}
The expressions for $f_i(r)$ function are given in \ref{appA}.

\section{Boundary conditions}
To show the continuity for the interior spacetime to the external one, one need boundary matching conditions. The exterior solution is given by Glavan and Lin \cite{glavan2020einstein}  in the limit $D\rightarrow 4$ as
\begin{eqnarray}
ds^2 &=&- F(r) dt^2 + {dr^2 \over F(r)} + r^2 d\Omega_2^2,
\end{eqnarray}
where,
\begin{eqnarray}
F(r) &=& 1+{r^2 \over 32\pi \alpha} \left[1 \pm \left\{1+{128\pi \alpha M \over r^3} \right\}^{1/2} \right]. \label{e27}
\end{eqnarray}

The above exterior spacetime has no meaningful solution at short distances if $\alpha<0$, however, $\alpha>0$ has two branches of solution. The asymptotic nature (at $r \rightarrow \infty$) of the two branches ( `$-$' or `$+$') for $\alpha>0$ are
\begin{eqnarray}
F(r)\rightarrow 1-{2M \over r}\rightarrow 1~~\mbox{or}~~1+{r^2 \over 16\pi \alpha}+{2M \over r} \rightarrow \infty.
\end{eqnarray}
Hence, the negative branch is asymptotically Minkowski's space while the positive branch blows up. Further, the nature of \eqref{e27} at $\alpha \rightarrow 0$ takes the form (`$-$' and `$+$ ' branches respectively)
\begin{eqnarray}
F(r)\rightarrow 1-{2M \over r} ~~\mbox{and} ~~1+{r^2 \over 16\pi \alpha}+{2M \over r} \rightarrow \infty.
\end{eqnarray}
This means that the negative branch reduces to Schwarzschild's vacuum while the positive branch blows up again as $\alpha \rightarrow 0$. Therefore, a negative branch of \eqref{e27} coincides with Schwarzschild exterior and further with Minkowski's spacetime asymptotically and also at $\alpha \rightarrow 0$. Hence, the negative branch is preferred in four-dimensional spacetime,
\begin{eqnarray}
F(r) &=& 1+{r^2 \over 32\pi \alpha} \left[1 - \left\{1+{128\pi \alpha M \over r^3} \right\}^{1/2} \right]. \label{e29}
\end{eqnarray}
Now matching the interior and exterior spacetime at the boundary $r=R$, we get
\begin{eqnarray}
e^{\nu(R)} = e^{-\lambda(R)}=F(R).
\end{eqnarray}
From this boundary condition, we get
\begingroup
\small
\begin{eqnarray}
a &=& \frac{1-\sqrt{128 \pi  \alpha  M/R^3+1}}{R^2 \left(\sqrt{128 \pi  \alpha  M/R^3+1}-1\right)-32 \pi  \alpha },\\
A &=& \ln \left[\frac{R^2 \left(1-\sqrt{128 \pi  \alpha  M/R^3+1}\right)}{32 \pi  \alpha }+1\right] - {1 \over 8} \times \nonumber \\
&& \hspace{0.0cm} \Bigg[ \frac{a \gamma  (a \alpha -1)}{\pi  \left(a R^2+1\right)^2} - \frac{4 a^2 \alpha  \gamma }{3 \pi  \left(a R^2+1\right)^3}+ \frac{16 \pi  \chi  \left(a R^2+1\right)^2}{a} \nonumber \\
&& \hspace{0.0cm}  -\frac{a \gamma  (a \alpha +8)}{4 \left(\pi  a R^2+\pi \right)} + 4 \left(a R^2+1\right) (1-16 \pi  \alpha  \chi +\beta)+ \nonumber \\
&& \hspace{0.0cm}  \frac{(64 \pi  \beta -5 a \gamma )\, \ln \left(a R^2+1\right)}{8 \pi }+\frac{a}{8 \pi }~ \ln \left(2 a \alpha +a R^2+1\right) \nonumber \\
&& \hspace{0.0cm}  \big\{1024 \pi ^2 \alpha ^2 \chi -96 \pi  \alpha  (\beta +1)+9 \gamma \big\}\Bigg].
\end{eqnarray}
\endgroup

And at the boundary, the radial pressure has to vanish i.e. $p_r(R)=0$ which gives
\begin{eqnarray}
\chi &=& -\frac{a \beta  \left[a^2 \left(R^4-\alpha  R^2\right)+a \left(3 \alpha +4 R^2\right)+3\right]}{8 \pi  \left(a R^2+1\right)^3}- \nonumber \\
&& \frac{a^2 \gamma  \left[a^2 \left(R^4-\alpha  R^2\right)+a \left(3 \alpha +4 R^2\right)+3\right]^2}{64 \pi ^2 \left(a R^2+1\right)^6}
\end{eqnarray}
\section{Physics analysis}
We will examine the physical properties of our current model in this section. The following aspects of EGB gravity theory must be checked for anisotropic neutron stars in order to achieve this goal.
\subsection{Metric potentials}
The metric potentials satisfy  $e^{\nu(0)}=$ constant and $e^{\lambda(0)}=1$. The gravitational metric potentials at the center of the stellar model's configuration are finite, according to the above calculation. Furthermore, at the center, the derivatives of these potentials are finite. The metric is regular in the center and performs nicely throughout the interior of the stellar model owing to the aforementioned requirements.
\subsection{Pressure and density}
The central density ($\rho_c$) and central radial pressure ($p_{rc}$) for our present model are obtained as,
\begin{eqnarray}
\rho_c &=& \frac{a (3 a \alpha +3)}{8 \pi } ,\\
p_{rc} &=& \gamma\, \rho_c^2+\beta \rho_c+\chi.
\end{eqnarray}

\begin{figure}
\centering
\includegraphics[width=8cm,height=6cm]{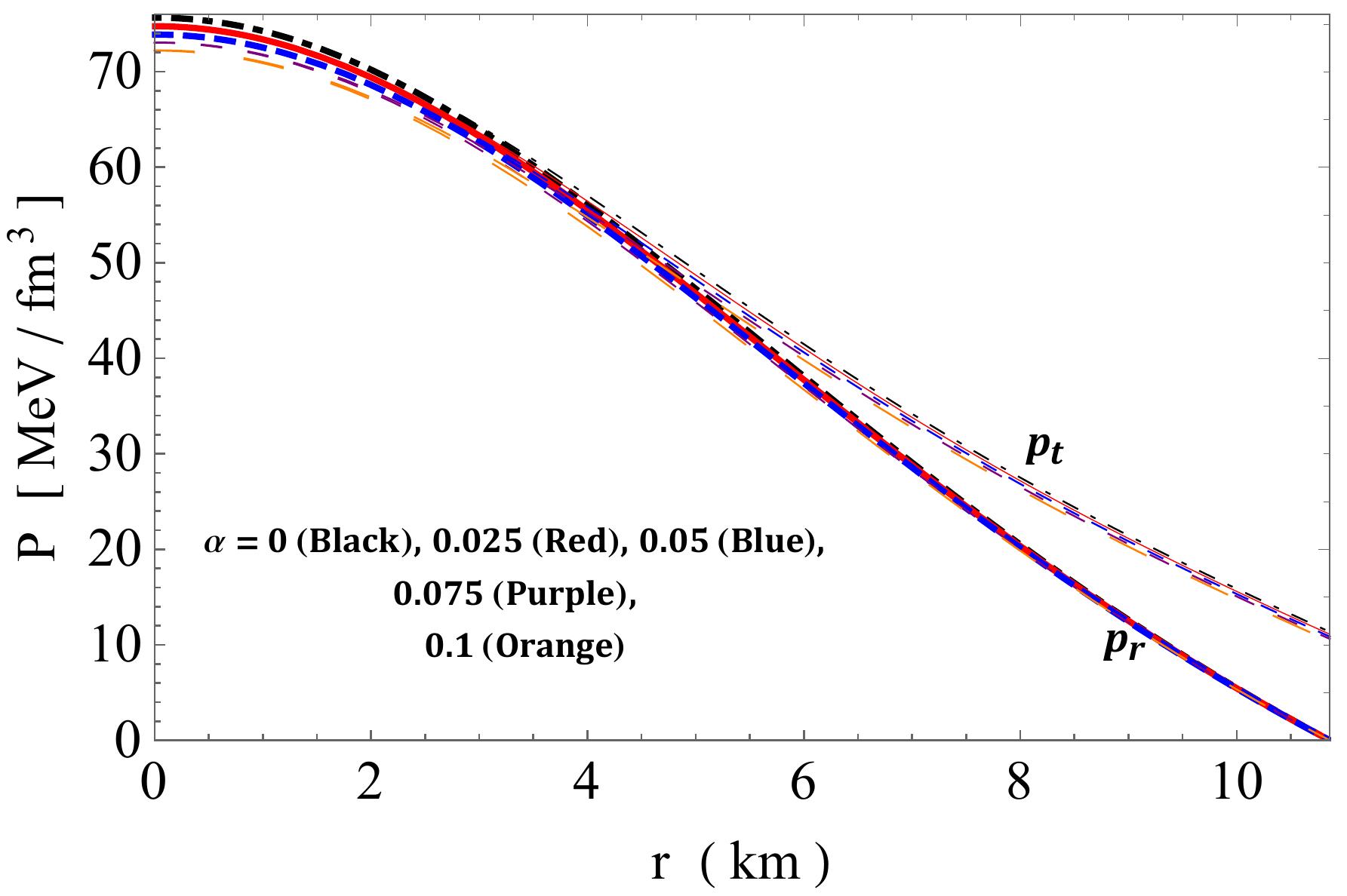}
\includegraphics[width=8cm,height=6cm]{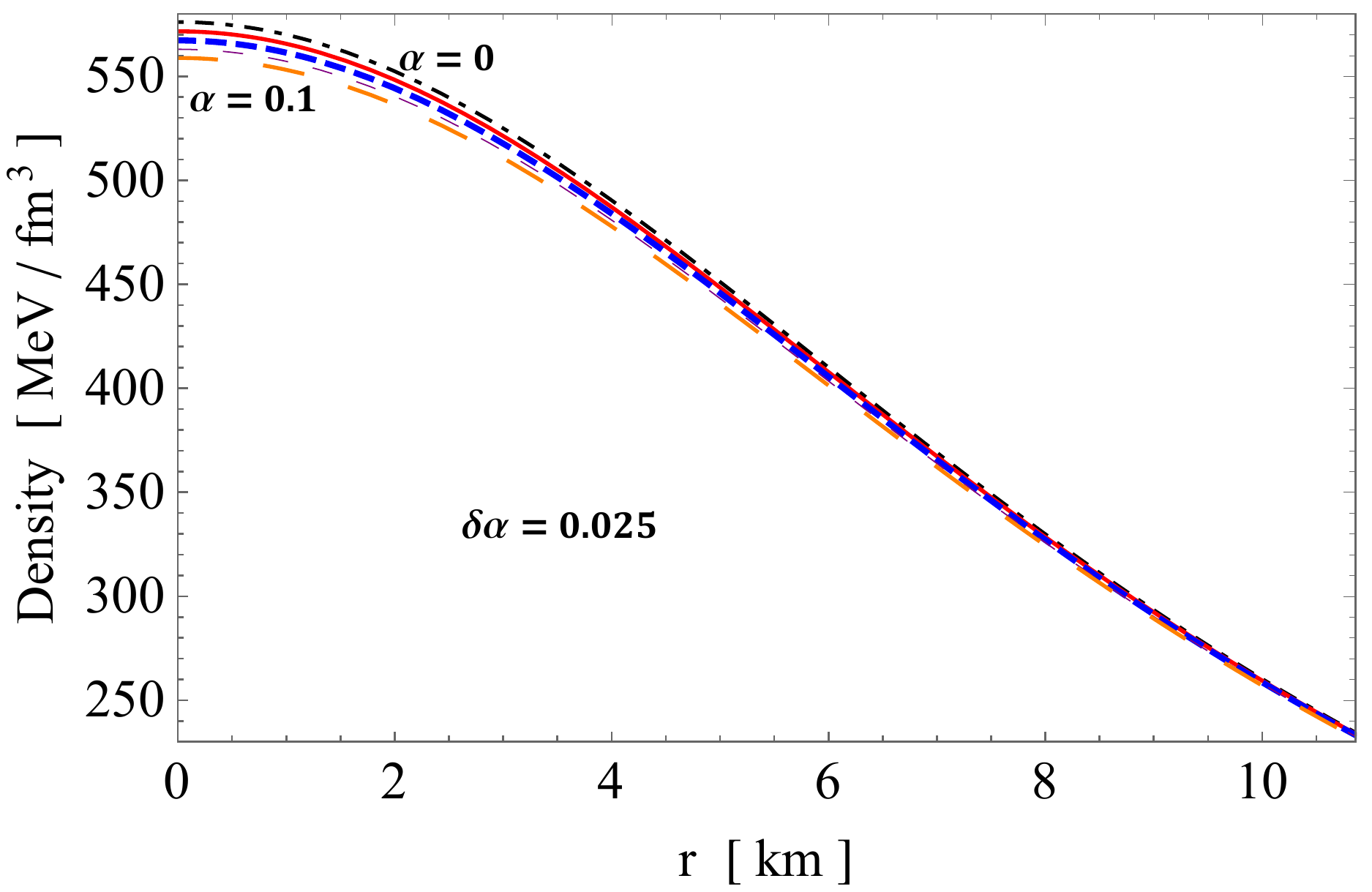}
\caption{The behavior of radial and tangential pressures (top) and energy density $(\rho)$ (bottom) versus radial coordinate $r$ for different values of $\alpha$ taking $M = 1.58M_\odot, R = 10.86km, \gamma = 20, \beta = 0.2$.}
\label{f1}
\end{figure}

One can note that both are finite inside the stellar interior and the central density has a linear dependence on $\alpha$ coupling.
Fig. \ref{f1} show that the density, radial, and tangential pressures are the monotonic decreasing function of `$r$'. Both pressures and density are positive and finite inside the stellar interior and hence physically acceptable.

\subsection{Causality condition and adiabatic index}
\begin{figure}
\includegraphics[width=8cm,height=6cm]{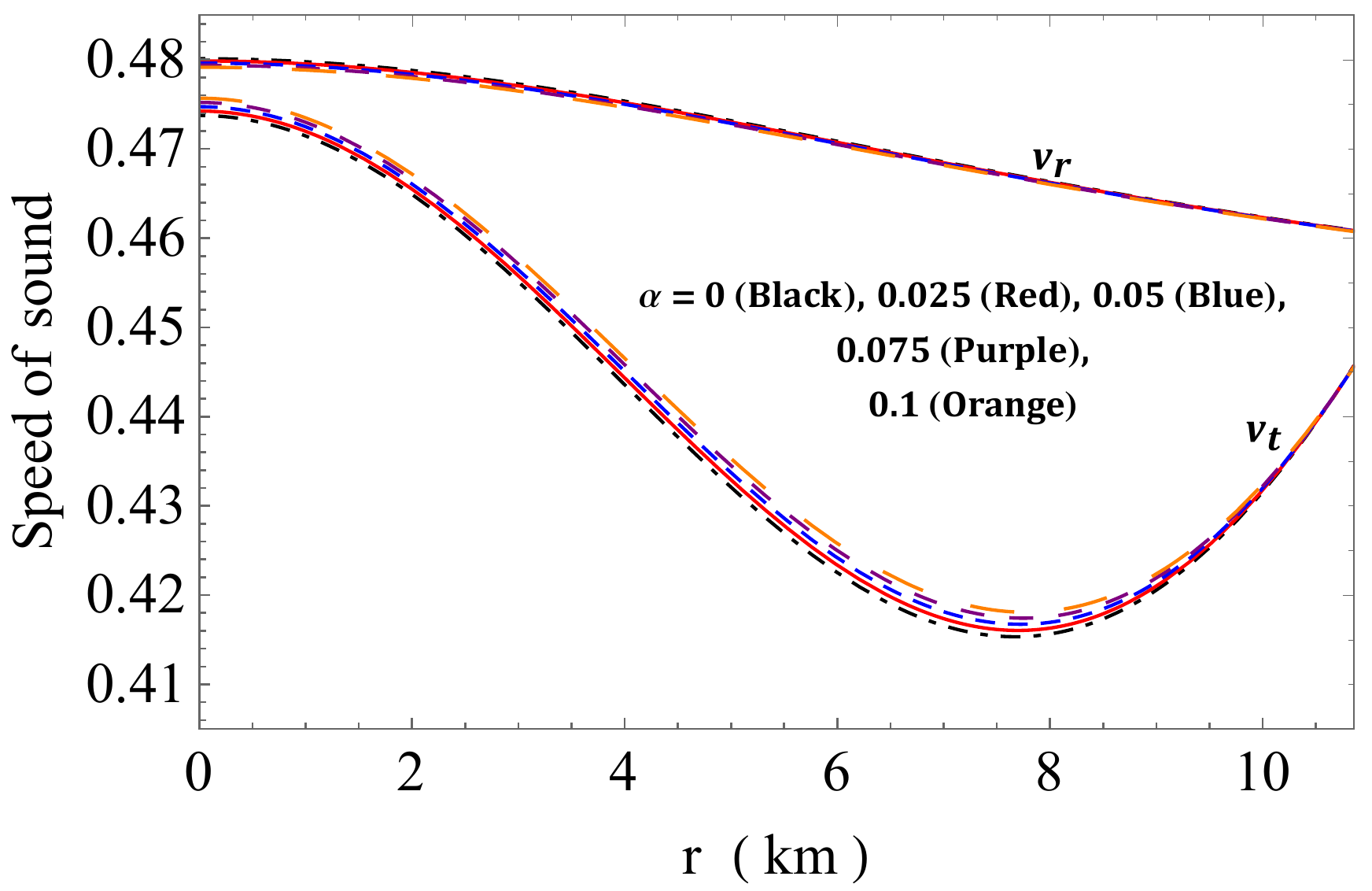}
\caption{The behavior of radial and tangential speed of sound ($v_r$ and $v_t$) versus radial coordinate $r$ for different values of $\alpha$ taking $M = 1.58M_\odot, R = 10.86km, \gamma = 20, \beta = 0.2$..}
\label{f4}
\end{figure}

\begin{figure}
\includegraphics[width=8cm,height=6cm]{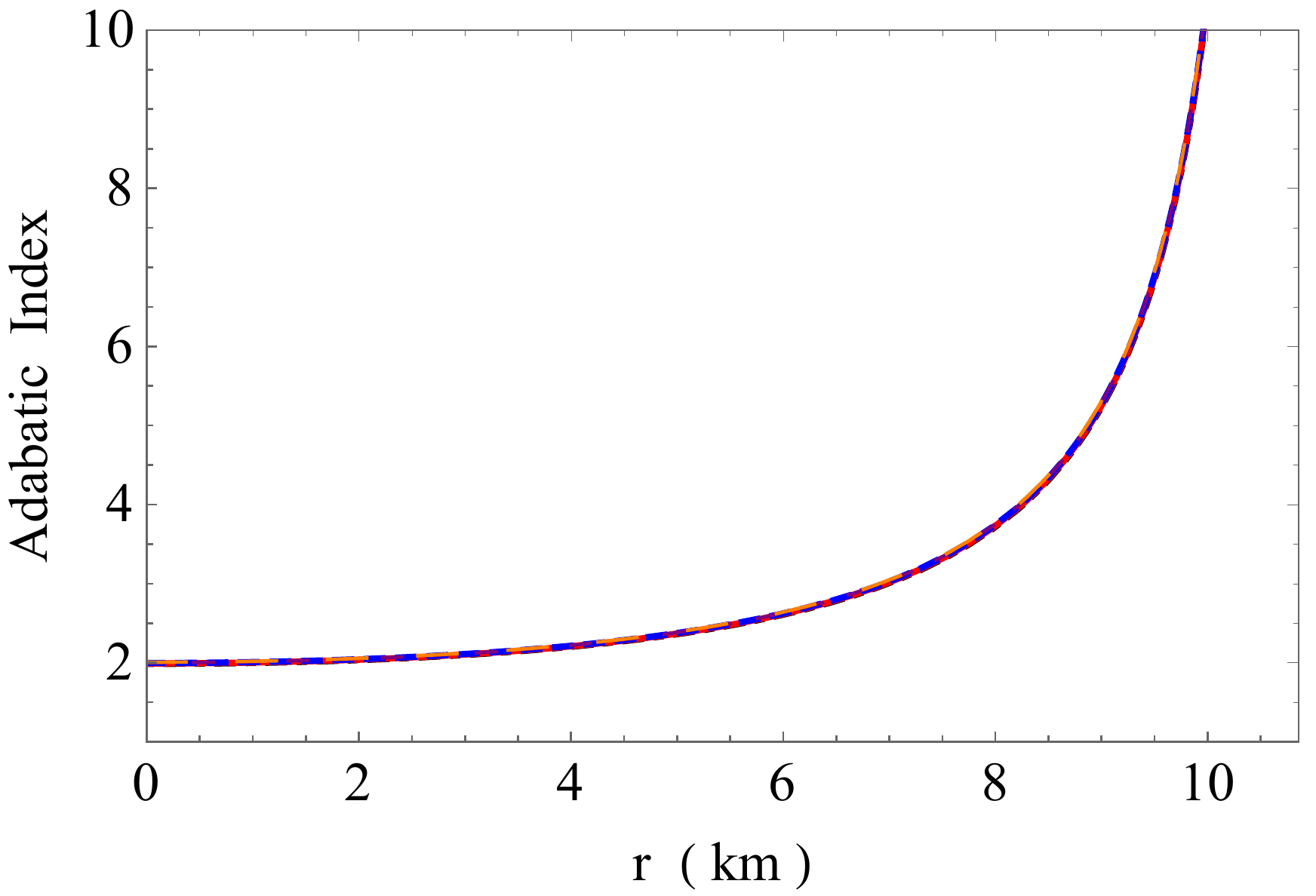}
\caption{The behavior of adiabatic index $(\Gamma)$ versus radial coordinate $r$ for different values of $\alpha$ taking $M = 1.58M_\odot, R = 10.86km, \gamma = 20, \beta = 0.2$.}
\label{f5}
\end{figure}

By considering the speed of sound, one can manage the stability analysis of compact objects. The rule that the speed of light exceeds the speed of sound is always followed by a physically acceptable solution which is termed a causality condition. We calculate the radial and transverse components of sound speed denoted by $v_r$ and $v_t$, respectively. These two components should be less than the speed of light. Fig. \ref{f4} clearly shows that the speeds of both sounds in radial and tangential directions obey the causality condition, ensuring the physical viability of the present solution. Abreu et al. proposed another stability condition based on these velocities in the literature \cite{ab1}, which is described as $-1<v_r^2-v_t^2<1$. This is considered to be one of the most interesting aspects of the neutron star model. We can confirm this condition using Fig. \ref{f4} that $v_r>v_t$ implying that the  Abreu et al. \cite{ab1} condition holds. Now, it is obvious that our model is consistent with this stability criterion as well. To discuss the stability of stellar configurations, the adiabatic index $\Gamma=\Big(1+\frac{\rho}{p_r}\Big) \frac{dp_r}{d\rho}$  is a useful tool which was Chandrasekhar for determining dynamical stability against infinitesimal radial adiabatic perturbation of the stellar system. But this study requires the critical adiabatic index $\Gamma_{cr}$, for the starting of instability, increases because of relativistic effects from the Newtonian value $\Gamma=4/3$. Therefore, in order to have a stable configuration against the radial perturbations, the value of $\Gamma$ must be greater than $\Gamma_{cr}$ and it is required to describe compact objects such as white dwarfs, neutron stars and supermassive stars \cite{Moustakidis}. On the other hand, Haensel et al. \cite{Haensel} has shown that the value of  $\Gamma$ lies between $2$ and $4$ for the EoS related to
neutron star matter. In this connection, the adiabatic index value for relativistic polytropic stars depending on the central value of pressure-density ratio was obtained by Glass \& Harpaz \cite{Glass} which is $\Gamma>4/3$. For our model, the value of the adiabatic index is greater than $4/3$ as can be seen in Fig. \ref{f5} that confirming the stability of our proposed model. 

\subsection{Pressure anisotropy}
\begin{figure}
\includegraphics[width=8cm,height=6cm]{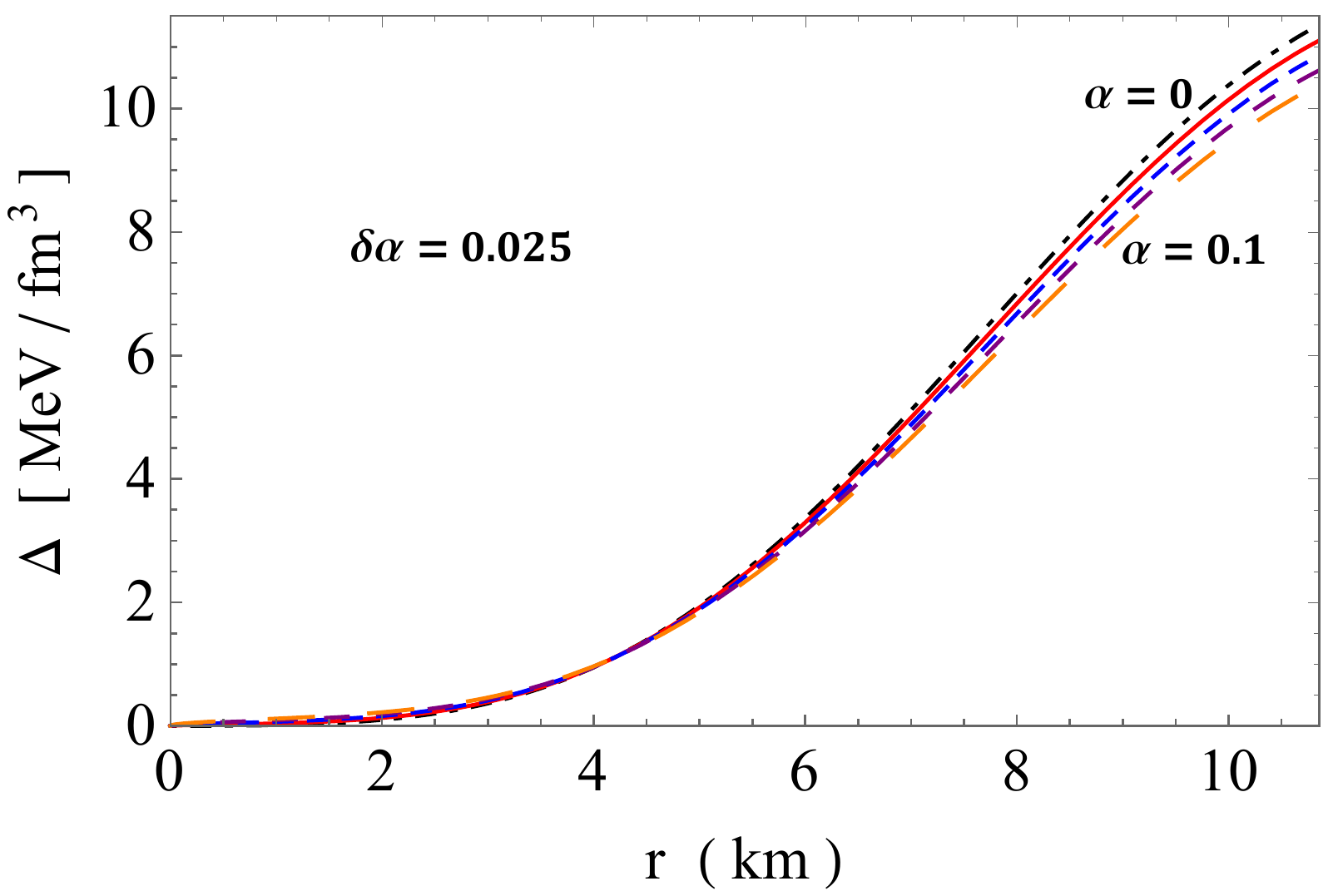}
\caption{The behavior of anisotropy $(\Delta)$ versus radial coordinate $r$ for different values of $\alpha$ taking $M = 1.58M_\odot, R = 10.86km, \gamma = 20, \beta = 0.2$.}
\label{f2}
\end{figure}

The internal structure of relativistic stellar objects can be illustrated by the term anisotropy in neutron star modeling, which offers information on the anisotropic behavior of the model. Fig. \ref{f2} shows the anisotropy behavior graphically. If $ p_t> p_r$, anisotropic pressure is directed outward, resulting in $\Delta>0$, whereas if $p_t <p_r$, anisotropy becomes negative, resulting in $\Delta<0$, indicating that anisotropic pressure is drawn inward. The graphical analysis of anisotropic measurement shows that for our proposed model $ p_t>p_r$ and hence the anisotropic is repulsive.

\subsection{Static stability criterion and gravitational redshift}
\begin{figure}
\includegraphics[width=8.5cm,height=6.3cm]{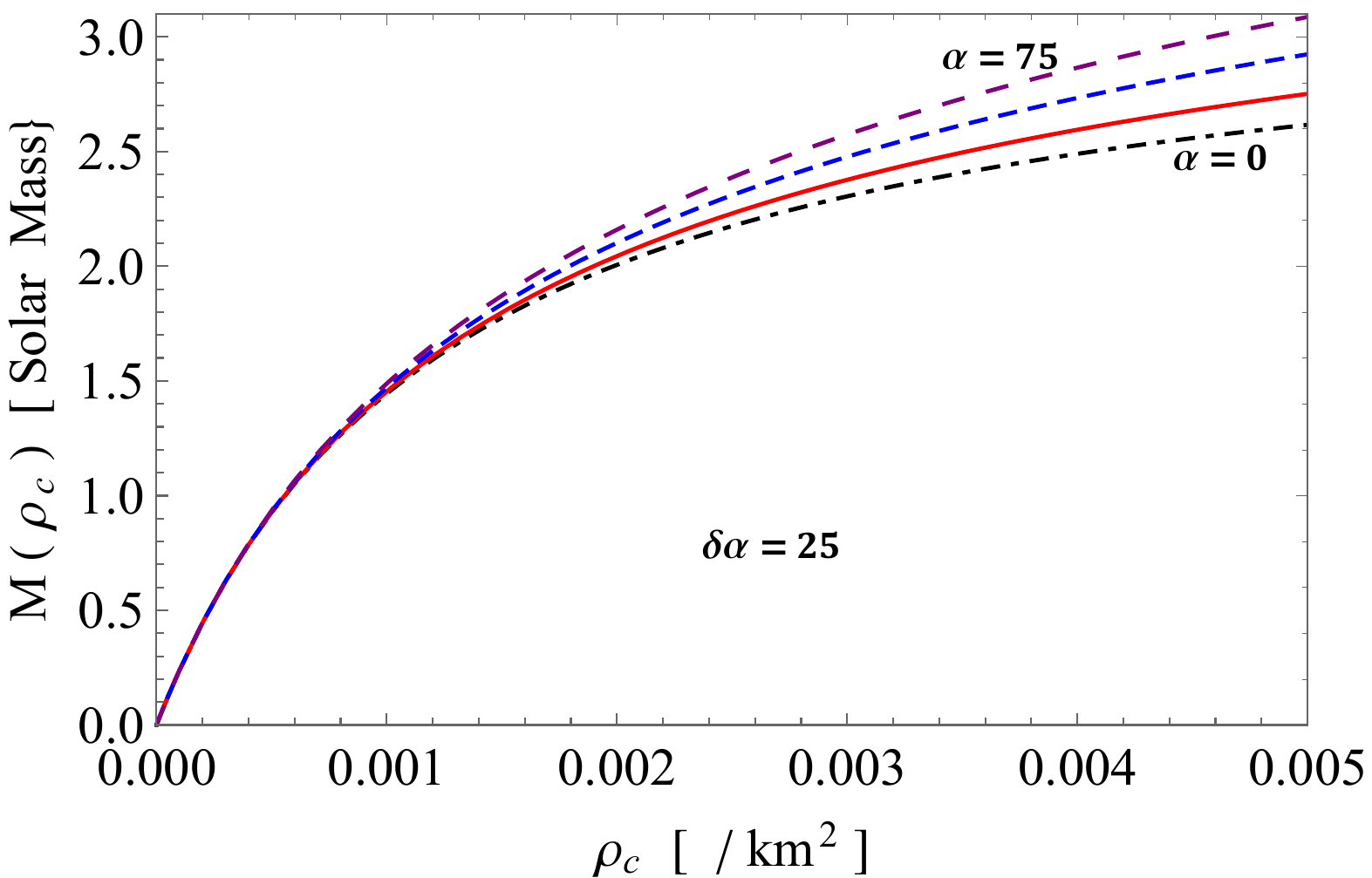}
\caption{$M-\rho_c$ curve for different values of $\alpha$.}
\label{f7}
\end{figure}

\begin{figure}
\includegraphics[width=8.5cm,height=6.4cm]{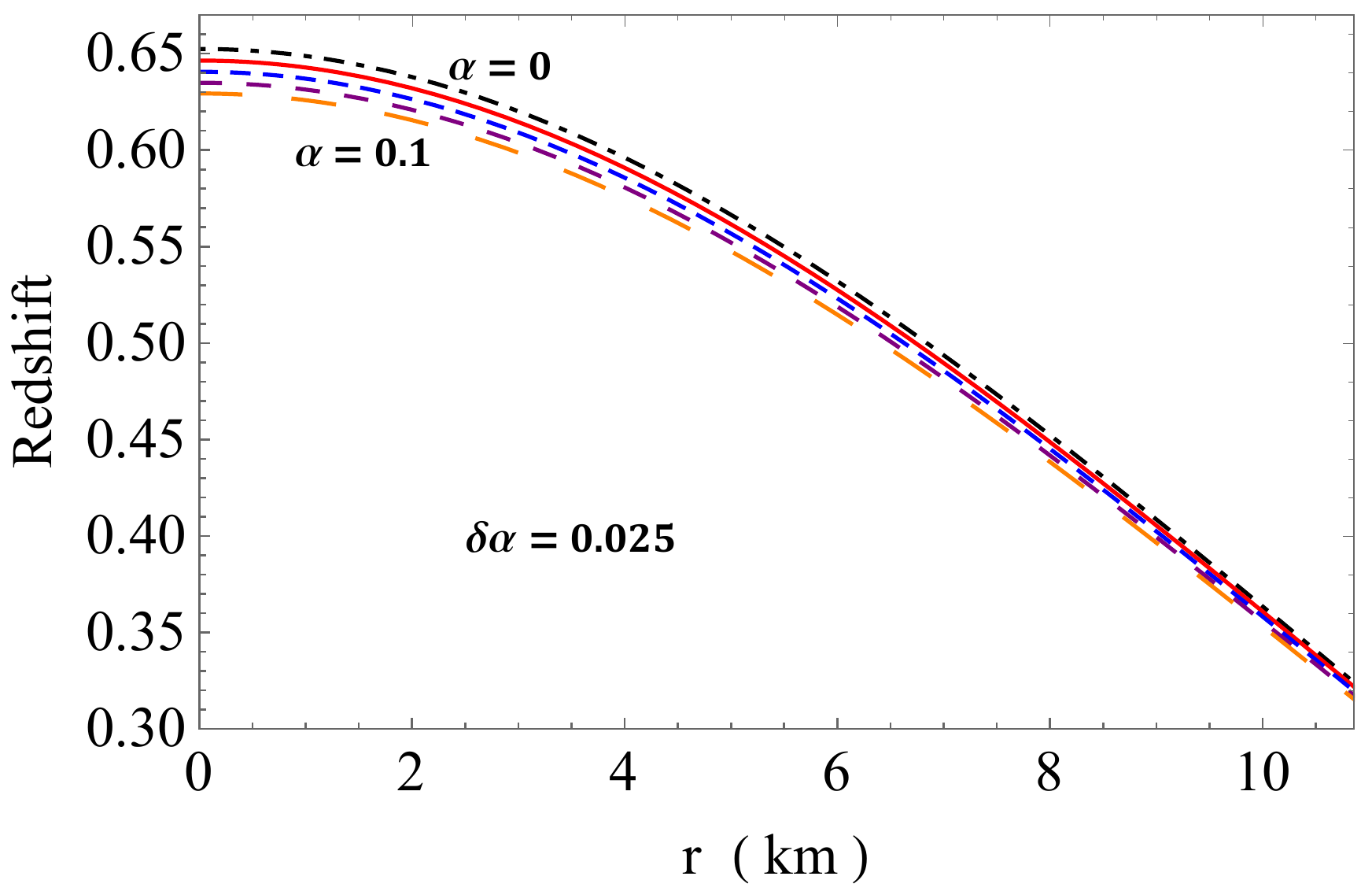}
\caption{The behavior of redshift curve versus radial coordinate $r$ for different values of $\alpha$ taking $M = 1.58M_\odot, R = 10.86km, \gamma = 20, \beta = 0.2$.}
\label{f3}
\end{figure}

For a neutron star to be stable under density perturbation, the mass of the system has to be an increasing function of its central density. The mass as a function of its central density ($\rho_c$) is found to be
\begin{small}
\begin{eqnarray}
\hspace{0.2cm} m(\rho_c)=\frac{R^5 \left(48 \pi  \alpha  \text{$\rho_c$}-3 \sqrt{96 \pi  \alpha  \rho_c+9}+9\right)+48 \pi  \alpha ^2 \rho_c R^3}{1.4766 \left[6 \alpha +R^2 \left(\sqrt{96 \pi  \alpha \rho_c+9}-3\right)\right]^2}.
\end{eqnarray}
\end{small}
The variation of mass with central density is given in Fig. \ref{f7}. It can be seen that the mass increases its central density implicating that the solution is stable. Further, one can also observe that for higher GB coupling $\alpha$ the configuration can hold more mass for a given range of density perturbation. This means that the $\alpha$ coupling enhances the stability of the neutron star. On the other hand, the gravitational redshift of the $4D-$EGB model is calculated by temporal component of the metric function as $z=\sqrt{e^{-\nu(r)}}-1$. It is found that the surface redshift value is decreasing when the coupling constant $\alpha$ is increasing (see Fig. \ref{f3}). The obtained values of the surface redshift values are: $z_s=0.65$ for $\alpha=0$, and $z_s=0.63$ for $\alpha=0.1$. 

\subsection{Predicted radii for some known compact star through $M-R$ diagram}
To incorporate the solution with the recent observational results, we have plotted and fitted a few well-known compact stars. Here, it can be seen that the maximum mass in the $M-R$ decreases with an increase in Gauss-Bonnet coupling strength. Further, we have fitted the observational data for three neutron stars so that their radii can be predicted from the $M-R$ curve. Their predicted radii are given in Table \ref{table2}. Further, to strengthen the discussions we have also incorporated the observations from GW170817 for which neutron stars of masses 1.6$M_\odot$ and $1.4 M_\odot$ must have radius at least $10.68^{+0.15}_{-0.04}~km$ \cite{bau} and $11.0^{+0.9}_{-0.6}~km$ \cite{cap} respectively. The neutron star of mass $1.6M_\odot$ is well fitted with $\alpha=1$ and $1.4M_\odot$ with $\alpha=0$ and $\alpha=1$. Lighter neutron star configurations will be well fitted with lower $\alpha$.

\begin{figure}[H]
\includegraphics[width=8cm,height=6cm]{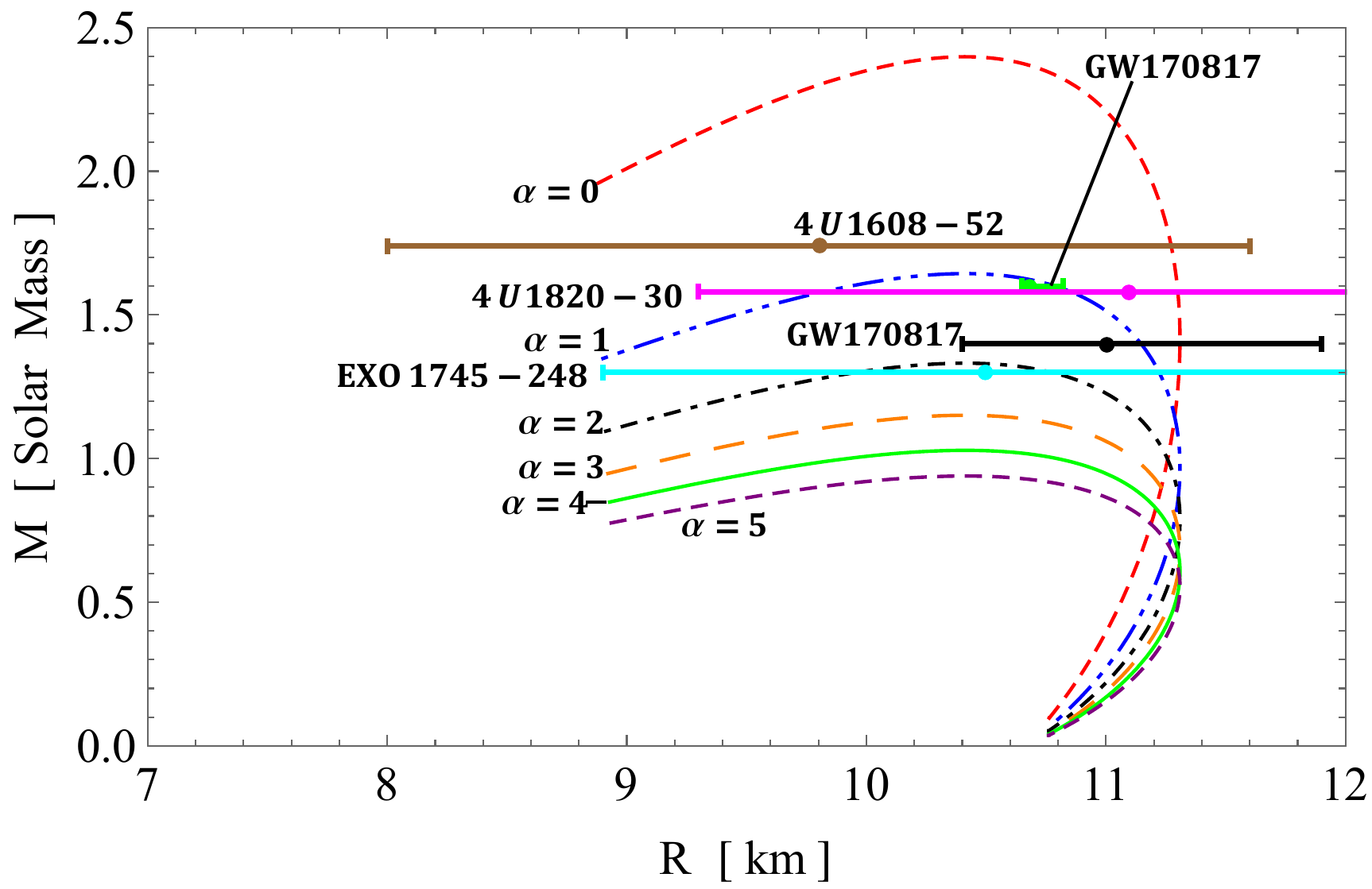}
\caption{$M-R$ curve for different values of $\alpha$.}
\label{f6}
\end{figure}

\begin{table}
\centering
\caption{ The predicted radii for observed compact star fitted by $M-R$ curve for different values of $\alpha$.}\label{Tab2}
\resizebox{1.03\hsize }{!}{$\begin{tabular}{|c|c|cccccc|}
\hline
{Compact stars}  &{Observed Mass } & \multicolumn{6}{|c|}{ Predicted Radii ($km$)}\\
\cline{3-8}
 & ($M_{\odot}$) & $\alpha=0$ &$\alpha=1$ & $\alpha=2$ &  $\alpha=3$ & $\alpha=4$ & $\alpha=5$ \\
\hline
4U 1608-52 & 1.74~\cite{guv}  &  11.27 & - & - & - & - & -  \\
4U 1820-30 & 1.58 ~\cite{guv1} & 11.31 & 10.86 & - & - & - & -  \\
EXO 1745-248 & 1.3 ~\cite{oz}  & 11.30 & 11.22 & 10.78 & - & - & -\\
\hline
GW170817 & 1.6 ~\cite{bau}  & 10.78 & - & - & - & - & -\\
GW170817 & 1.4 ~\cite{cap}  & 11.31 & 11.15 & - & - & - & -\\
\hline
\end{tabular}$}\label{table2}
\end{table}

\begin{figure}[H]
\includegraphics[width=8.8cm,height=7cm]{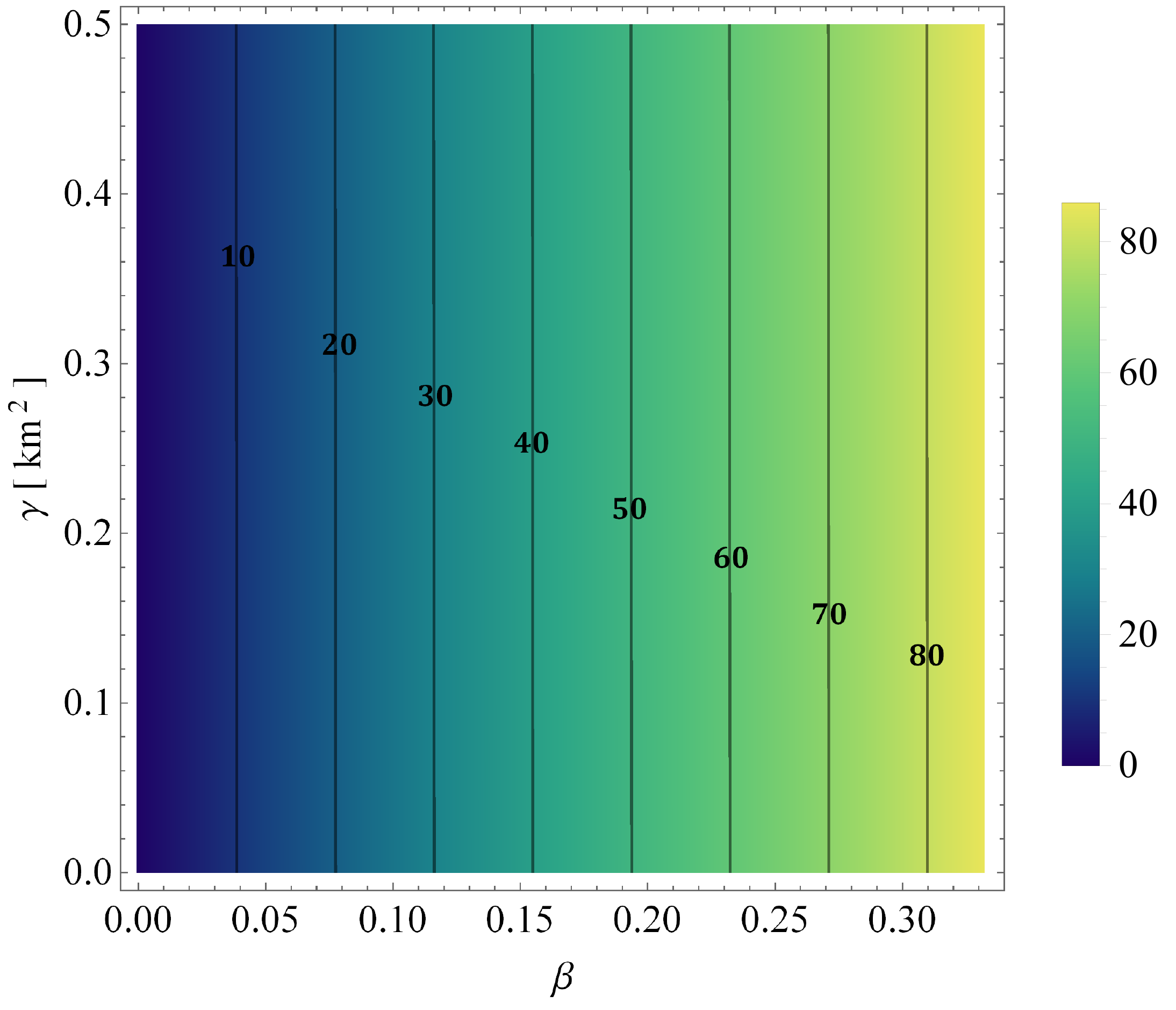}\\
\includegraphics[width=8.8cm,height=7cm]{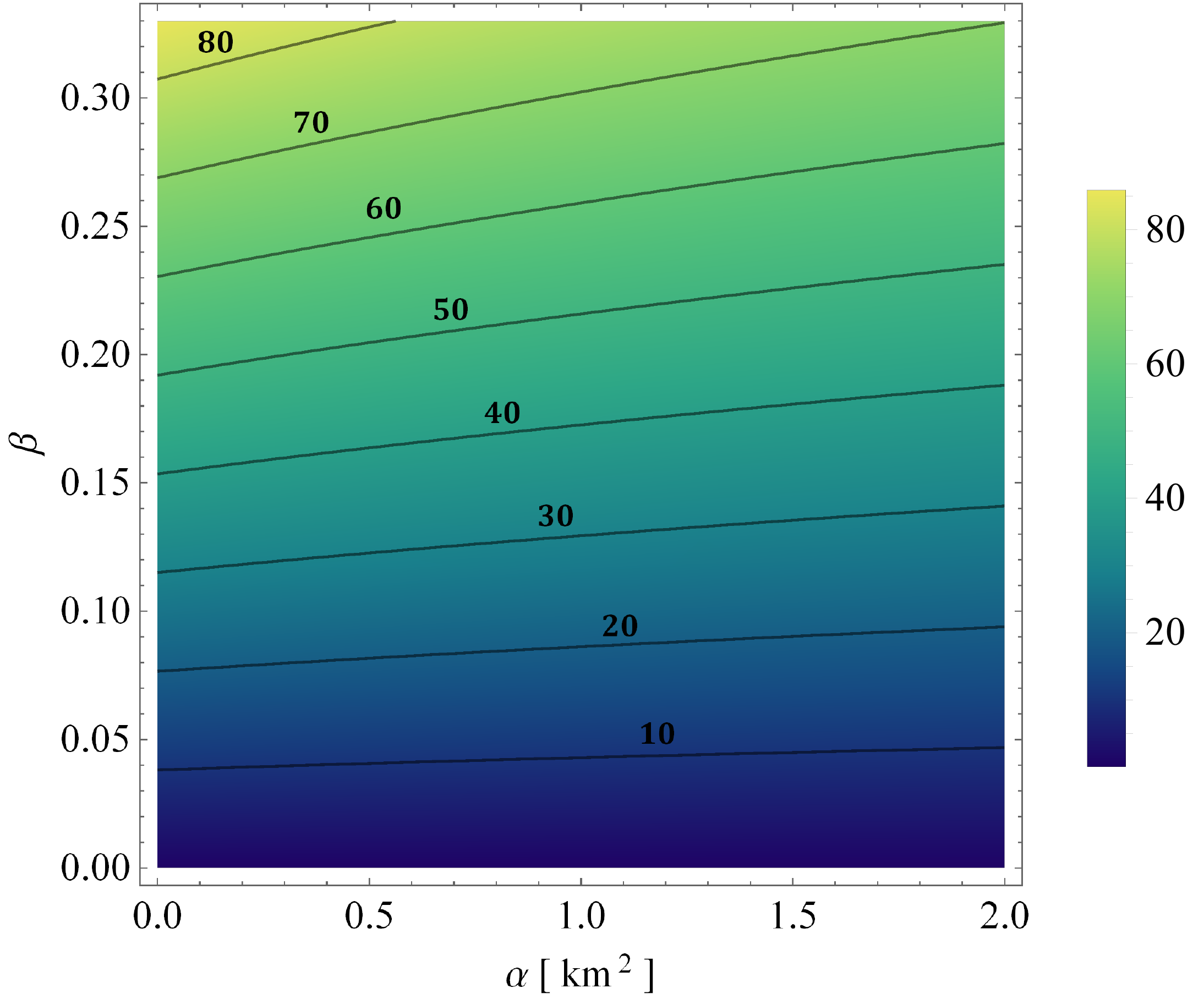}
\caption{$\gamma-\beta$ (top) and $\beta-\alpha$ (bottom) planes for equi-$\mathcal{B}_g$ contours .}
\label{f8}
\end{figure}

\subsection{Mass and Bag constants measurements of anisotropic star models through $equi-plane$ diagram}
\begin{figure}[H]
\includegraphics[width=8.8cm,height=7cm]{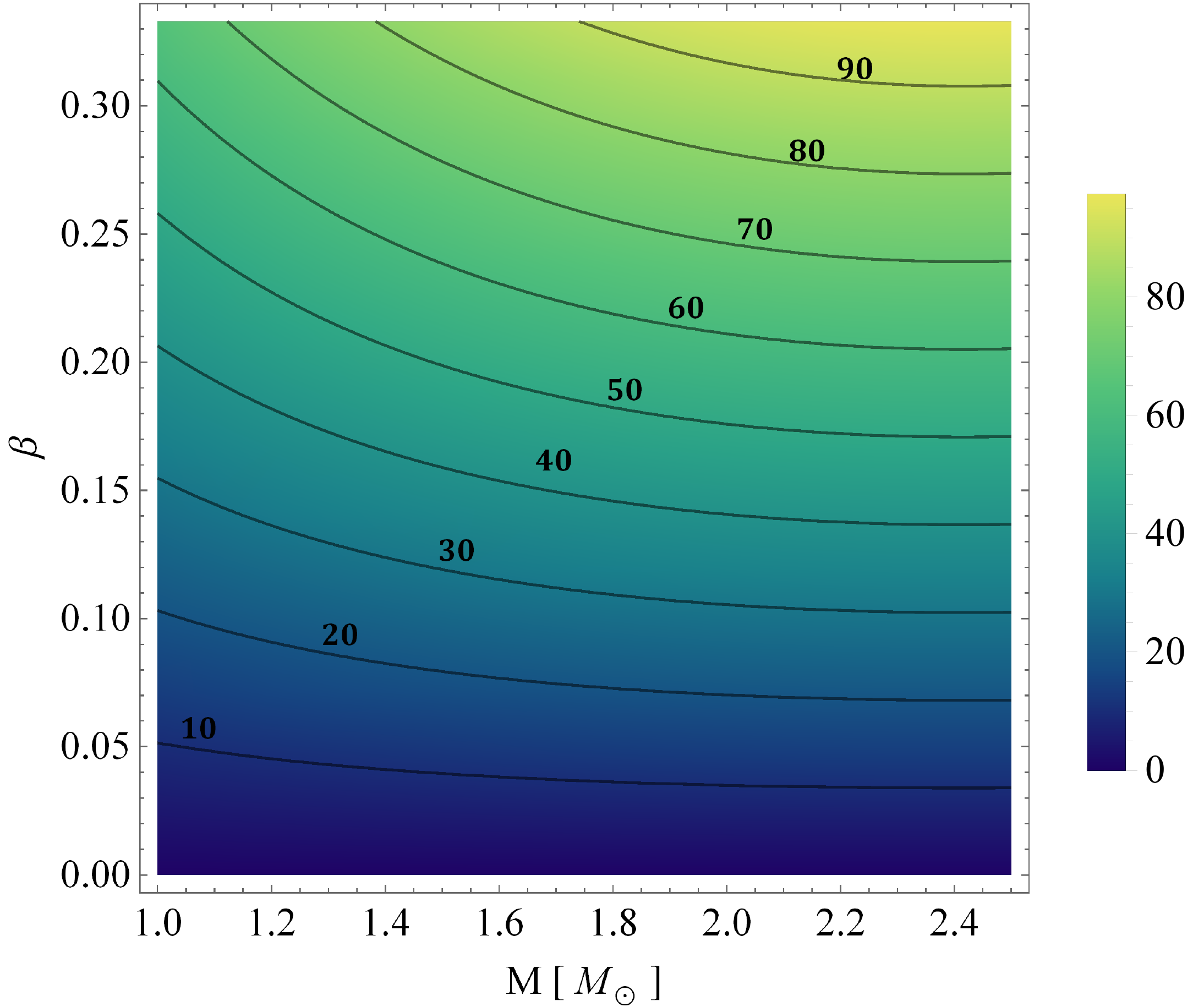}\\
\includegraphics[width=8.8cm,height=7cm]{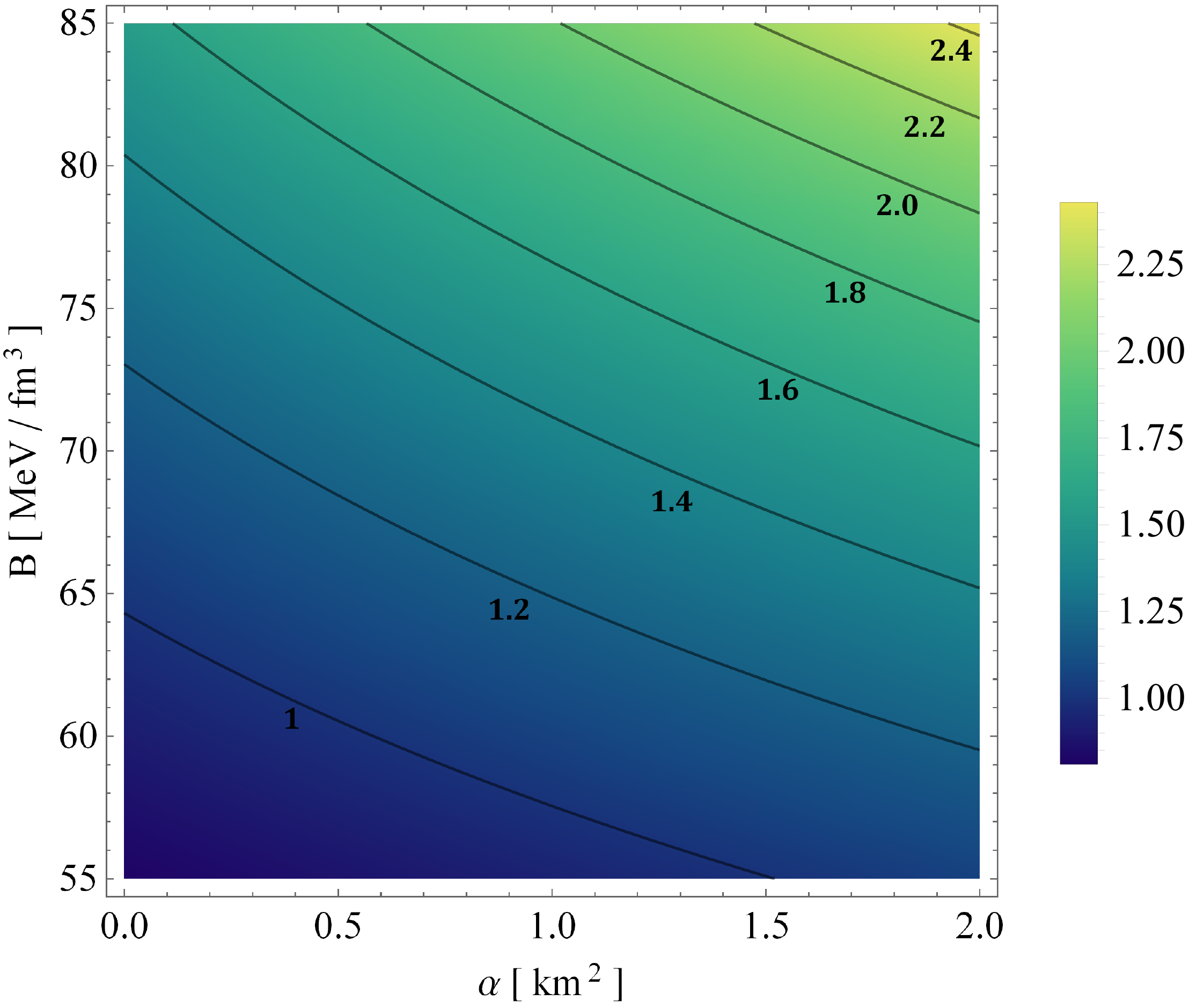}
\caption{$\beta-M$ planes for equi-$\mathcal{B}_g$ contour (top) and $\alpha-B$ (bottom) plane for equi-$M$ contour.}
\label{f9}
\end{figure}

For more detailed analysis, a few plane graphs were plotted. The equi-$\mathcal{B}_g$ contours are plotted in the $\beta-\gamma$ plane (Fig. \ref{f8} top panel) and $\alpha-\beta$ plane (Fig. \ref{f8} bottom panel). It can be seen from the top panel of Fig. \ref{f8}  that as the value of $\beta$ increases, the bag constant $\mathcal{B}_g$ increases, while it doesn't show any change with the variation of $\gamma$. Whereas in the bottom panel of Fig. \ref{f8} we can see that for the $\alpha-\beta$ plane, the bag constant $\mathcal{B}_g$ slightly increases with the value of $\alpha$. For both these plots, the Bag constant stays within the expected range ($40 - 120 MeV / fm^3$) in the short interval $0.15 \le \beta \le 0.35$. Now, the $equi-\mathcal{B}_g$ contours are plotted for $M-\beta$ plane in Fig. \ref{f9} top panel. As we can see here, with the increase in mass, the value of $\mathcal{B}_g$ decreases, and the decrease becomes sharper for higher values of $\beta$. In Fig. \ref{f9} bottom panel, the $equi-mass$ contour is plotted for $\alpha-\mathcal{B}_g$ plane and it can be seen that the mass increases with the increase of both $\mathcal{B}_g$ and the Gauss-Bonnet constant $\alpha$. Therefore, we can conclude that the supermassive solution is favourable at a higher bag constant ($\mathcal{B}_g$) and high $\alpha$.

\begin{figure}[H]
\includegraphics[width=8.3cm,height=6cm]{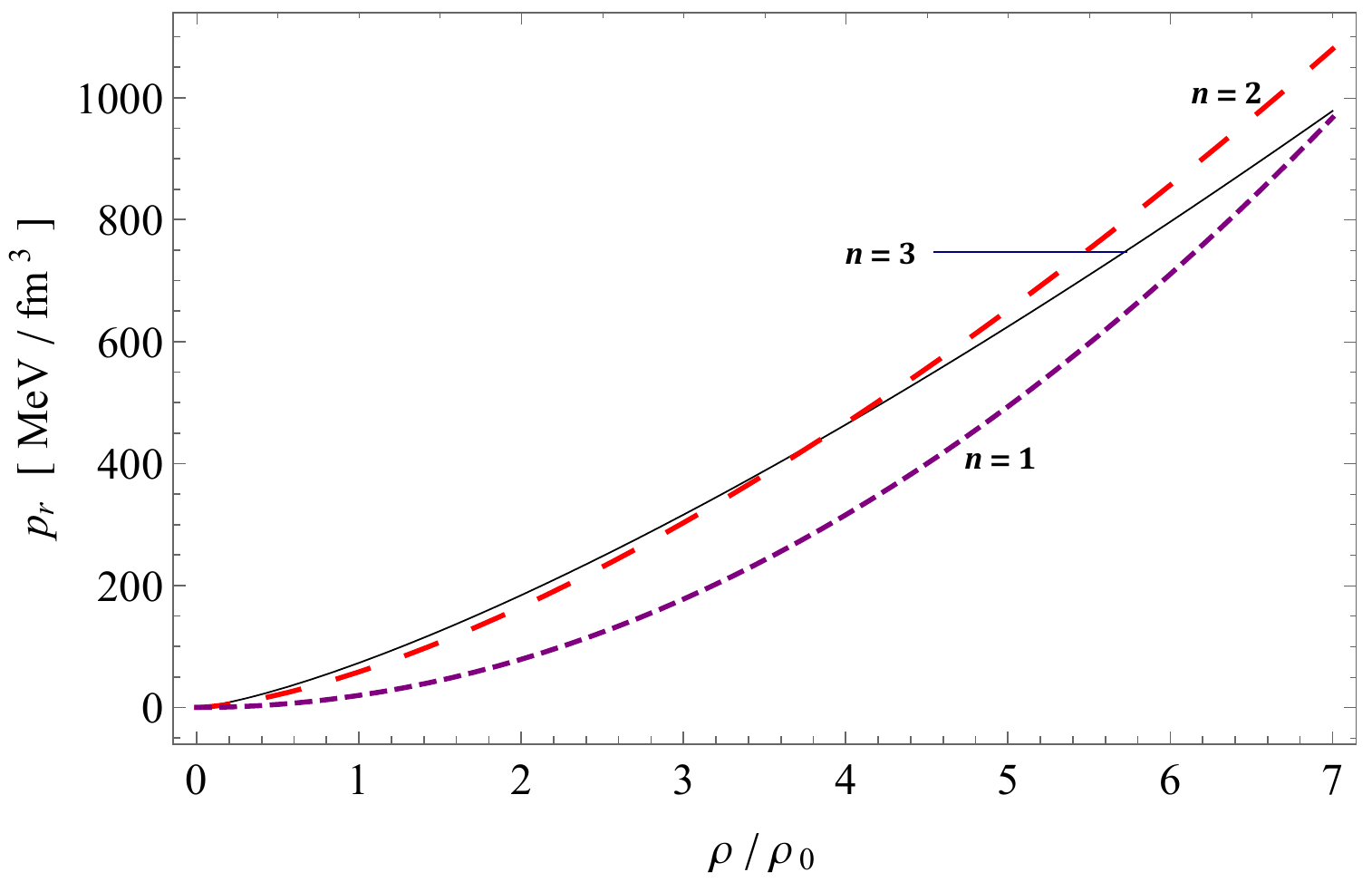}
\caption{Equation of State for different values of $n$-parameter. Here the density is measured in the unit of nuclear saturation density $\rho_0=2.5\times 10^{14}g/cm^3$.}
\label{f10}
\end{figure}

\section{Results and discussion}
We have successfully obtained a new exact solution in four-dimensional EGB gravity satisfying a polytropic equation of state. This solution fulfills all the physical criteria such as causality condition, Bondi criterion, Abreu et al. condition, and static stability criterion. Figure \ref{f1} shows the variations of pressure and density with radius. Here we can see that the central pressure and density decrease with an increase in $\alpha$. In a similar way, Fig. \ref{f2} shows the variation of anisotropy where the surface anisotropy decreases with an increase in $\alpha$. The satisfaction of the causality condition can be seen in Fig. \ref{f4}. Further, one can also see that the stability factor $v_t^2-v_r^2$ is always negative within the stellar interior signifying that the solution is stable under anisotropy perturbation. Figure \ref{f5} shows the variation of adiabatic index with radius.  

The central values of the adiabatic index ($\Gamma_c \sim 1.95$) have very minor changes when the GB-coupling changes. The comparison of redshift between GR ($\alpha=0$) and EGB ($\alpha \neq 0$) is shown in Fig. \ref{f3} where both the central and surface redshifts in EBG gravity are always less than that of GR counterpart.  The mass-radius relationship is shown via $M-R$ curve in Fig. \ref{f7}. This curve is in well agreement with the recent gravitational wave observation from GW 170817 where for a neutron stars of masses 1.6$M_\odot$ and $1.4 M_\odot$ must have radius at least $10.68^{+0.15}_{-0.04}~km$ \cite{bau} and $11.0^{+0.9}_{-0.6}~km$ \cite{cap} respectively. From our $M-R$ curve, these neutron stars i.e. $1.6M_\odot$ and $1.4M_\odot$ are well-fitted with $\alpha=0$ and $\alpha=0,1$ curves respectively, and their corresponding predicted radii are 10.78 $km$ and (11.31 $km$, 11.15 $km$) respectively. Hence, the solution is in agreement with the observations from the neutron star merger GW170817. Further, it can be noted that the bag constant in this polytropic extension on MIT EoS is independent of the polytropic parameter $\gamma$, however, slightly increases with increase in Gauss-Bonnet coupling strength. Again, more massive compact structures are favourable with higher bag constant and higher $\beta-$parameter (see Figs. \ref{f8} and \ref{f9}). The equation of state is plotted for different values of the polytropic parameter $n$ in Fig. \ref{f10}. Here we can see that the quadratic contribution is more at the lower density regime. As the density increases the linear contribution increases behaving more like MIT bag EoS. This implies that at the higher density regime the nucleon matter de-confines into asymptotically free quarks.

Finally, one can conclude that the proposed anisotropic solution is satisfying all the physical and mathematical requirements which represent realistic celestial bodies at least from a theoretical point of view.

\section*{Acknowledgement}
The author acknowledges Prof. Sushant Ghosh, Jamia Millia Islamia, New Delhi for helping us in deriving the field equations. The author SKM acknowledges that this work is carried out under TRC Project (Grant No. BFP/RGP/CBS-/19/099), the Sultanate of Oman. SKM is thankful for continuous support and encouragement from the administration of University of Nizwa. 

\appendix
\section{Expressions for the functions $f_i(r)$}\label{appA}

\begin{footnotesize}
\begin{eqnarray}
f_1(r) &=& \frac{32 \alpha  \gamma  a^3}{\pi  \left(a r^2+1\right)^4}+\frac{2 \pi  (a \alpha +8) \gamma  a^2}{\left(a \pi  r^2+\pi \right)^2}-\frac{16 (a \alpha -1) \gamma  a^2}{\pi  \left(a r^2+1\right)^3}+ \nonumber \\
&& \frac{(64 \pi  \beta -5 a \gamma ) a}{a \pi  r^2+\pi } \\
f_2(r) &=& \frac{\left(1024 \pi ^2 \chi  \alpha ^2-96 \pi  (\beta +1) \alpha +9 \gamma \right) a^2}{a \pi  \left(r^2+2 \alpha \right)+\pi }+32 (\beta -16 \pi  \alpha  \chi +1)  \nonumber \\
&& a+256 \pi  \left(a r^2+1\right) \chi \\
f_3(r) &=& \frac{32 \alpha  \gamma  a^3}{\pi  \left(a r^2+1\right)^4}+\frac{2 \pi  (a \alpha +8) \gamma  a^2}{\left(a \pi  r^2+\pi \right)^2}-\frac{16 (a \alpha -1) \gamma  a^2}{\pi  \left(a r^2+1\right)^3}+ \nonumber \\
&& \frac{(64 \pi  \beta -5 a \gamma ) a}{a \pi  r^2+\pi }-\frac{64 a}{a r^2+1} \\
f_4(r) &=& -\frac{16 r^2 \alpha  \gamma  a^4}{\pi  \left(a r^2+1\right)^5}+\frac{2 \alpha  \gamma  a^3}{\pi  \left(a r^2+1\right)^4}+\frac{6 r^2 (a \alpha -1) \gamma  a^3}{\pi  \left(a r^2+1\right)^4} \nonumber \\
&& -\frac{(a \alpha -1) \gamma  a^2}{\pi  \left(a r^2+1\right)^3} \\
f_5(r) &=& -\frac{r^2 (a \alpha +8) \gamma  a^3}{2 \pi  \left(a r^2+1\right)^3}+\frac{\pi  (a \alpha +8) \gamma  a^2}{8 \left(a \pi  r^2+\pi \right)^2}+\frac{r^2 (5 a \gamma -64 \pi  \beta ) a^2}{8 \pi  \left(a r^2+1\right)^2}-\nonumber \\
&& \frac{(5 a \gamma -64 \pi  \beta ) a}{16 \left(a \pi  r^2+\pi \right)}\\ 
f_6(r) &=& -\frac{r^2 \left(1024 \pi ^2 \chi  \alpha ^2-96 \pi  (\beta +1) \alpha +9 \gamma \right) a^3}{8 \pi  \left(a \left(r^2+2 \alpha \right)+1\right)^2}+32 \pi  r^2 \chi  a+\nonumber \\
&& \frac{\left(1024 \pi ^2 \chi  \alpha ^2-96 \pi  (\beta +1) \alpha +9 \gamma \right) a^2}{16 \left(a \pi  \left(r^2+2 \alpha \right)+\pi \right)}+2 (\beta -16 \pi  \alpha  \chi +1) \nonumber \\
&& a+16 \pi  \left(a r^2+1\right) \chi.
\end{eqnarray}
\end{footnotesize}

\section{Regularization of EGB theory in $4D$}\label{appB}
The $D-$dimensional EGB field equations for the spacetime \eqref{e9} before re-scaling are given by
\begin{footnotesize}
\begin{eqnarray}
8\pi \rho &=& {(D-2)\lambda'e^{-\lambda} \over 2r}\Big[1+{2\hat{\alpha}(D-3)(D-4) \over r^2}  \big(1-e^{-\lambda}\big)\Big]+ \nonumber \\
&& {(D-2)(1-e^{-\lambda}) \over 2r^2} \Big[(D-3)+{\hat{\alpha}(D-3)(D-4)(D-5) \over r^2} \nonumber \\
&& \big(1-e^{-\lambda}\big) \Big],\\
8\pi p_r &=& {(D-2)\nu'e^{-\lambda} \over 2r}\Big[1+{2\hat{\alpha}(D-3)(D-4) \over r^2}  \big(1-e^{-\lambda}\big)\Big]- \nonumber \\
&& {(D-2)(1-e^{-\lambda}) \over 2r^2} \Big[(D-3)+{\hat{\alpha}(D-3)(D-4)(D-5) \over r^2} \nonumber \\
&& \big(1-e^{-\lambda}\big) \Big],\\
8\pi p_t &=& {e^{-\lambda} \over 4} \Big[\big(2\nu''+\nu'^2 \big) \Big\{1+{4\hat{\alpha}(D-3)(D-4) \over r^2}  \big(1-e^{-\lambda}\big) \Big\} +\nonumber \\
&& {2(\nu'-\lambda') \over r} \Big\{(D-3) +{2\hat{\alpha}(D-3)(D-4)(D-5) \over r^2}  \big(1-e^{-\lambda}\big) \Big\} \nonumber \\
&& -\nu' \lambda' \Big\{1+{8\hat{\alpha}(D-3)(D-4)(D-5) \over r^2}  + {12\hat{\alpha}(D-3)(D-4) \over r^2} \nonumber \\
&& \big(1-e^{-\lambda}\big)\Big\} \Big]-{1-e^{-\lambda} \over r^2} \Big\{(D-3)(D-4)+ \big(1-e^{-\lambda}\big) \nonumber \\
&& {\hat{\alpha}(D-3)(D-4)(D-5)(D-6) \over r^2}\Big\}.
\end{eqnarray}
\end{footnotesize}
After the re-scaling $\hat{\alpha} \rightarrow \alpha/(D-4)$, we get
\begin{footnotesize}
\begin{eqnarray}
8\pi \rho &=& {(D-2)\lambda'e^{-\lambda} \over 2r}\Big[1+{2\alpha(D-3) \over r^2}  \big(1-e^{-\lambda}\big)\Big]+ \nonumber \\
&& {(D-2)(1-e^{-\lambda}) \over 2r^2} \Big[(D-3)+{\alpha(D-3)(D-5) \over r^2} \nonumber \\
&& \big(1-e^{-\lambda}\big) \Big]
\end{eqnarray}
\begin{eqnarray}
8\pi p_r &=& {(D-2)\nu'e^{-\lambda} \over 2r}\Big[1+{2\alpha(D-3) \over r^2}  \big(1-e^{-\lambda}\big)\Big]- \nonumber \\
&& {(D-2)(1-e^{-\lambda}) \over 2r^2} \Big[(D-3)+{\alpha(D-3)(D-5) \over r^2} \nonumber \\
&& \big(1-e^{-\lambda}\big) \Big]\\
8\pi p_t &=& {e^{-\lambda} \over 4} \Big[\big(2\nu''+\nu'^2 \big) \Big\{1+{4\alpha(D-3) \over r^2}  \big(1-e^{-\lambda}\big) \Big\} +\nonumber \\
&& {2(\nu'-\lambda') \over r} \Big\{(D-3) +{2\alpha(D-3)(D-5) \over r^2}  \big(1-e^{-\lambda}\big) \Big\} \nonumber \\
&& -\nu' \lambda' \Big\{1+{8\alpha(D-3)(D-5) \over r^2}  + {12\alpha(D-3) \over r^2} \nonumber \\
&& \big(1-e^{-\lambda}\big)\Big\} \Big]-{1-e^{-\lambda} \over r^2} \Big\{(D-3)(D-4)+ \big(1-e^{-\lambda}\big) \nonumber \\
&& {\alpha(D-3)(D-5)(D-6) \over r^2}\Big\}.
\end{eqnarray}
\end{footnotesize}
Finally, we can take $D\rightarrow 4$ and we get
\begin{footnotesize}
\begin{eqnarray}
8\pi \rho &=& {\lambda'e^{-\lambda} \over r}\bigg[1+{2\alpha\big(1-e^{-\lambda}\big) \over r^2} \bigg]+ {1-e^{-\lambda} \over r^2} \bigg[1-{\alpha \big(1-e^{-\lambda}\big) \over r^2} \bigg] \nonumber\\
8\pi p_r &=& {\nu'e^{-\lambda} \over r}\bigg[1+{2\alpha \big(1-e^{-\lambda}\big) \over r^2}  \bigg]- {1-e^{-\lambda} \over r^2} \bigg[1-{\alpha \big(1-e^{-\lambda}\big) \over r^2} \bigg] \nonumber\\
8\pi p_t &=& {e^{-\lambda} \over 4} \bigg[\big(2\nu''+\nu'^2 \big) \bigg\{1+{4\alpha \big(1-e^{-\lambda}\big) \over r^2} \bigg\} + {2(\nu'-\lambda') \over r} \nonumber \\
&&  \bigg\{1 -{2\alpha \big(1-e^{-\lambda}\big) \over r^2}   \bigg\}  -\nu' \lambda' \bigg\{1-{8\alpha \over r^2}  + {12\alpha \big(1-e^{-\lambda}\big) \over r^2} \bigg\} \bigg]\nonumber \\
&& -{2\alpha \left(1-e^{-\lambda} \right)^2 \over r^4}, \nonumber
\end{eqnarray}
\end{footnotesize}
which is the required field equations in $4D-$EGB gravity.


\begin{thebibliography}{99}
\bibitem{Lovelock1} D. Lovelock, J. Math. Phys. {\bf 13}, 874 (1972).
\bibitem{Lovelock2}  D. Lovelock, J. Math. Phys. {\bf 12}, 498 (1972).
\bibitem{Tomozawa}  Y. Tomozawa, arXiv:1107.1424.
\bibitem{Glavan} D. Glavan , C. Lin,  Phys. Rev. Lett. {\bf 124}, 081301 (2020).
\bibitem{Doneva1} D.D. Doneva, S.S. Yazadjiev, arXiv:2003.10284.
\bibitem{g8} S.G. Ghosh, R. Kumar, Class. Quantum Gravity {\bf 37}, 245008 (2020).
\bibitem{g9} R.A. Konoplya, A. Zhidenko, Phys. Dark Univ. {\bf 30}, 100697 (2020) .
\bibitem{g10} D.V. Singh, S. Siwach, Phys. Lett. B {\bf 808}, 135658 (2020).
\bibitem{g11} S.A.H. Mansoori, Phys. Dark Univ. {\bf 31}, 100776 (2021).
\bibitem{g12} D.V. Singh, S.G. Ghosh, S.D. Maharaj, Phys. Dark Univ. {\bf 30}, 100730 (2020).
\bibitem{g13} S.W. Wei, Y.X. Liu, Phys. Rev. D {\bf 101}, 104018 (2020).
\bibitem{g14} K. Yang, B.M. Gu, S.W. Wei, Y.X. Liu, Eur. Phys. J. C {\bf 80}, 662 (2020)
\bibitem{g15} P.G.S. Fernandes, Phys. Lett. B {\bf 805}, 135468 (2020).
\bibitem{g16} C.Y. Zhang, S.J. Zhang, P.C. Li, M. Guo, JHEP {\bf 2008}, 105 (2020).
\bibitem{g17} K. Jusufi, Ann. Phys. {\bf 421}, 168285 (2020).
\bibitem{g18} A. Abdujabbarov, J. Rayimbaev, B. Turimov, F. Atamurotov, Phys. Dark Univ. {\bf 30}, 100715 (2020).
\bibitem{g19} K. Jafarzade, M. Kord Zangeneh, F.S.N.Lobo, arXiv:2009.12988 [gr-qc]
\bibitem{ruderman}R. Ruderman, Annu. Rev. Astron. Astrophys. {\bf 10}, 427 (1972).
\bibitem{Mak} M.K. Mak, T. Harko, Proc. R. Soc. Lond. A {\bf 459}, 393 (2003). 
\bibitem{Kipp} R.K. Kippenhahm, A. Weigert, Stellar Structure and Evolution (Springer, Berlin, 1990), p. 384
\bibitem{Soko} A.I. Sokolov, JETP {\bf 79},  1137 (1980).
\bibitem{Herrera1997}  L. Herrera, N.O. Santos, Phys. Rep. {\bf 286}, 53 (1997).
\bibitem{Hrr1} L. Herrera, V. Varela, Phys. Lett. A {\bf 189}, 11 (1994).
\bibitem{Hrr2} L. Herrera, W. Barreto, Phys. Rev. D {\bf 87}, 087303 (2013).
\bibitem{Hrr3} L. Herrera, W. Barreto, Phys. Rev. D {\bf 88}, 084022 (2013).
\bibitem{Hrr4} L. Herrera, A. Di Prisco, W. Barreto, and J. Ospino, Gen. Relativ. Gravit. {\bf 46}, 1827 (2014).
\bibitem{Sulaksono2019}A. Sulaksono, A. M. Setiawan, Eur. Phys. J. C {\bf 79},  755 (2019).
\bibitem{Rizaldy2019}  R. Rizaldy, A.R. Alfarasy, A. Sulaksono, T. Sumaryada, Phys. Rev. D {\bf 100}, 055804 (2019).
\bibitem{Maurya2019a}  S.K. Maurya, A. Banerjee, M.K. Jasim, J. Kumar, A.K. Prasad, A. Pradan, Phys. Rev. D {\bf 99}, 044029 (2019).
\bibitem{banerjee}A. Banerjee, T. Tangphati, D. Samart, P. Channuie, Astrophys. J. {\bf 906},  114 (2021).
\bibitem{banerjee1} A. Banerjee, T. Tangphati, D. Samart, P. Channuie, Astrophys. J. {\bf 909}, 114 (2021).
\bibitem{4D1} K. N. Singh, S. K. Maurya, A. Dutta, F. Rahaman, S. Aktar, Eur. Phys. J. C  {\bf 81}, 909 (2021).
\bibitem{4D2} K. Jusufi, A. Banerjee, S. G. Ghosh, Eur. Phys. J. C {\bf 80}, 698 (2020).
\bibitem{4D3} H. Sudan, A. Banerjee, L. Moodly, M. K. Jasim, Class. Quantum Grav. {\bf 38}, 035002 (2021). 
\bibitem{4D4}  T. Tangphati, A. Pradhan, A. Banerjee, G. Panotopoulos, Phys. Dark Univ. {\bf 33}, 100877 (2021).
\bibitem{lp} J. Lattimer, M. Prakash, Ap. J.  {\bf 550}, 426 (2001).
\bibitem{1974a}  A. Chodos,  R. L. Jaffe, K. Johnson,  C. B. Thorn, Phys. Rev. D {\bf 10}, 2599 (1974).
\bibitem{1974b} A. Chodos,  R. L. Jaffe, K. Johnson,  C. B. Thorn, V. Weisskopf, Phys. Rev. D {\bf 9}, 3471 (1974b),
\bibitem{2000} A. Peshier, B Kampfer,G. Soff, Phys. Rev. C {\bf 61}, 045203 (2000).
\bibitem{1999}  M. G. Alford, K. Rajagopal, F. Wilczek,   Nuc. Phys. B {\bf 537},  443 (1999).
\bibitem{1998} M. G. Alford, K. Rajagopal, F. Wilczek, Phys. Lett. B {\bf 422},  247 (1998).
\bibitem{2017} J. Asbell,  P. Jaikumar,  J. Phys. Conf. Ser. {\bf 861}, 012029 (2017)
\bibitem{Eos1} M.A. Abramowicz, Acta Astron. {\bf 33}, 313 (1983).
\bibitem{Eos2} M. Cosenza, L. Herrera, M. Esculpi, L. Witten, J. Math. Phys. {\bf 22}, 118 (1981).
\bibitem{Eos3} L. Herrera, W. Barreto, Gen. Relat. Gravity {\bf 36}, 127 (2004).
\bibitem{Eos4} L. Herrera et al., Phys. Rev. D {\bf 69}, 084026 (2004).
\bibitem{Eos5} L. Herrera, W. Barreto, Phys. Rev. D {\bf 88}, 084022 (2013).
\bibitem{Eos6} L. Herrera, A. Di Prisco, W. Barreto, J. Ospino, Gen. Relat. Gravity {\bf 46}, 1827 (2014).
\bibitem{Eos7} L. Herrera, E. Fuenmayor, P. Leon, Phys. Rev.D {\bf 93}, 024247 (2016).
\bibitem{Eos8} J.D. Bekenstein, Phys. Rev.D {\bf 4}, 2185 (1960).

\bibitem{Eos9} P.M. Takisa, S.D. Maharaj, Astrophys. Space Sci. {\bf 45}, 1951 (2013).
\bibitem{Eos14} M. Azam, S.A. Mardan JCAP {\bf 01}, 040 (2017). 
\bibitem{Eos1c} P.H. Chavanis, Eur. Phys. J. Plus {\bf 129}, 38 (2014).
\bibitem{Eos2c} P.H. Chavanis, Eur. Phys. J. Plus {\bf 129}, 222 (2014).
\bibitem{Eos3c} R.C. Freitas, S.V.B. Goncalves, Eur. Phys. J. C {\bf 74}, 3217 (2014).

\bibitem{Eos10} M. Azam, S.A. Mardan, M.A. Rehman, Astrophys. Space Sci. {\bf 359}, 14 (2015).
\bibitem{Eos11} R. Naeem, M. Azam, G. Abbas, H. Nazar, New Astronomy {\bf 89}, 101651 (2021).

\bibitem{Eos12} M. Azam, S.A. Mardan,  Eur. Phys. J. C {\bf 77}, 113 (2017).
\bibitem{Eos13}M. Azam, S.A. Mardan, I. Noureen et al., Eur. Phys. J. C {\bf 76}, 315 (2016).

\bibitem{glavan2020einstein} D. Glavan , C. Lin.  Phys. Rev. Lett. \textbf{124},  081301 (2020). 
\bibitem{ghm} S. G. Ghosh, S. D. Maharaj, Phys. Dark Univ. \textbf{30}, 100687 (2020). 
\bibitem{ab1}H. Abreu, H. Hernandez, L.A. Nunez, Class. Quantum Grav. {\bf 24}, 4631  (2007).

\bibitem{Moustakidis} C. C. Moustakidis, Gen. Relati. Gravi. {\bf 49},  68 (2017).

\bibitem{Haensel} P. Haensel, A. Y. Potekhin, D. G. Yakovlev, 2007, Neutron Stars 1: Equation of State and Structure (Berlin: Springer)

\bibitem{Glass} E. N. Glass, A. Harpaz,  MNRAS {\bf 202}, 1 (1983).

\bibitem{bau} A. Bauswein, et al., Astrophys. J. Lett. {\bf 850},  L34 (2017).
\bibitem{cap} C.D. Capano, et al., Nat. Astron. {\bf 4},  625 (2020).
\bibitem{guv} T. G$\Ddot{\text{u}}$ver et al.,  Ap. J. {\bf 719}, 1807 (2010).
\bibitem{guv1} T. G$\Ddot{\text{u}}$ver et al.,  Ap. J. {\bf 712}, 964 (2010).
\bibitem{oz}F. $\Ddot{\text{O}}$zel  Ap. J.  {\bf 693},  1775 (2009).

\bibitem{Charmousis} C. Charmousis et al., arXiv:2109.01149.
\bibitem{ban1}J. Pretel, A. Benerjee, arXiv:2107.03859.
\bibitem{clif}T. Clifton et al., Phys. Rev. D {\bf 102}, 084005 (2020).
\bibitem{feng}J. Feng et al., Phys. Rev. D {\bf 103}, 064002 (2021).
\bibitem{bhar1}P. Bhar et al., Eur. Phys. J. C  {\bf 77}, 109 (2017).
\bibitem{bhar2}P. Bhar, M. Govender,  Astrophys. Space Sci. {\bf 364}, 186 (2019).
\bibitem{bhar3} P. Bhar, et al., Eur. Phys. J. C {\bf 79}, 922 (2019).
\bibitem{Maurya1} S. K. Maurya et al.  Mod. Phys. Lett. A {\bf 36}, 2150231 (2021). 
\bibitem{Pedro} P. G. S. Fernandes et al., Class. Quantum Grav. 3{\bf 9}, 063001 (2022).


\end{thebibliography}

\end{document}